Dual Degree Project

# INVESTIGATIONS ON A THERMOACOUSTIC REFRIGERATOR

Submitted in partial fulfillment of the requirements for the degrees of

**Bachelor of Technology & Master of Technology**

by

**Ram Chandrashekhar Dhuley**
Roll No. 05D1----

Supervisor

**Prof. M. D. Atrey**

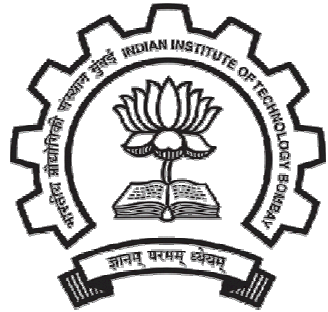

**Department of Mechanical Engineering**

**Indian Institute of Technology Bombay**
**Powai, Mumbai 400076**

**2010**

# APPROVAL SHEET

The thesis entitled **'Investigations on a Thermoacoustic Refrigerator'** by **Ram C. Dhuley** is approved for the degrees of Bachelor of Technology and Master of Technology.

| SUPERVISOR | EXAMINERS |
|---|---|
| | |
| **Prof. Milind D. Atrey** <br> (Dept. of Mech. Engg., IITB) | **Shri. Trilok Singh** <br> (BARC, Mumbai) |
| | |
| | **Prof. Amit Agrawal** <br> (Dept. of Mech. Engg., IITB) |
| | **CHAIRMAN** |
| | **Prof. Abhiram G. Ranade** <br> (Dept. of CSE, IITB) |

June 28, 2010

IIT Bombay, Mumbai

# Declaration

I declare that this written submission represents my ideas in my own words and where others' ideas or words have been included, I have adequately cited and referenced the original sources. I also declare that I have adhered to all principles of academic honesty and integrity and have not misrepresented or fabricated or falsified any idea/data/fact/source in my submission. I understand that any violation of the above will be cause for disciplinary action by the Institute and can also evoke penal action from the sources which have thus not been properly cited or from whom proper permission has not been taken when needed.

_________________________
(Signature)

*Ram C. Dhuley*
_________________________
(Name of the student)

_________________________
(Roll No.)

Date: 28/06/2010.

# ABSTRACT

Thermoacoustic Refrigerators use acoustic power for generating cold temperatures. Development of refrigerators based on thermoacoustic technology is a novel solution to the present day need of cooling, without causing environmental hazards. With added advantages like minimal moving parts and absence of CFC refrigerants, these devices can attain very low temperatures maintaining a compact size. The present work describes an in-depth theoretical analysis of standing wave thermoacoustic refrigerators. This consists of detailed parametric studies, transient state analysis and a design using available simulation software. Design and construction of a thermoacoustic refrigerator using a commercially available electro-dynamic motor is also presented.

# CONTENTS

# LIST OF FIGURES        PAGE

# LIST OF TABLES  PAGE

# NOMENCLATURE

| Symbol | Meaning (SI Unit) |
|---|---|
| p | pressure (N m$^{-2}$) |
| T | temperature (K) |
| f | frequency (Hz) |
| ρ | density (kg m$^{-3}$) |
| a | sound speed (m s$^{-1}$) |
| λ | wavelength (m) |
| k | wave number (m$^{-1}$) |
| ω | angular frequency (rad s$^{-1}$) |
| γ | ratio of specific heats |
| $δ_k$ | thermal penetration depth of gas(m) |
| $δ_s$ | thermal penetration depth of plate (m) |
| $δ_v$ | viscousl penetration depth of gas(m) |
| β | thermal expansion coefficient (k$^{-1}$) |
| k | thermal conductivity (W m$^{-1}$ k$^{-1}$) |
| $C_p$ | isobaric specific heat (J kg$^{-1}$ k$^{-1}$) |
| Γ | normalized temperature gradient |
| $f$ | Rott's function |
| s | specific entropy (J kg$^{-1}$ k$^{-1}$) |
| $\dot{Q}_c$ | cooling power (W) |
| $\dot{W}$ | acoustic power (W) |
| COP | coefficient of performance |
| L | length of resonator (m) |
| $L_s$ | length of stack (m) |
| $\varepsilon_s$ | Thermal Effusivity |
| l | half plate thickness (m) |
| $y_0$ | half plate spacing (m) |
| $\Pi$ | wetted perimeter (m) |
| $x_s$ | stack centre position (m) |
| x | Local x-coordinate (m) |
| A | Area of cross section (m$^2$) |

**Subscripts**

| | |
|---|---|
| m | mean |
| a | amplitude |
| 1 | local amplitude |
| s | Solid (plate) |
| n | normalized |
| r, res | resonator |

**Other Symbols**

| | |
|---|---|
| ~ | Complex Conjugate |
| Re[ ] | Real Part of [ ] |
| Im[ ] | Imaginary Part of [ ] |

# Chapter 1

# INTRODUCTION

## 1.1 Thermoacoustics

*Thermoacoustics* is the interaction between heat and sound. It explains how energy in form of heat can be converted to sound or how sound waves can be used to generate cold temperatures. The pressure and displacement oscillations in a sound wave are accompanied by temperature oscillations. For an adiabatic sound wave propagating through an ideal gas, the temperature oscillations, $T_1$ are related to the pressure oscillations $p_1$, [1] as:

$$\frac{T_1}{T_m} = \frac{\gamma-1}{\gamma} \frac{p_1}{p_m} \tag{1.1}$$

where $T_m$ and $p_m$ respectively, are the mean temperature and pressure of the medium, and $\gamma$ is the specific heat capacity ratio. In medium like air at STP and pressure amplitude of ordinary conversation (~ 60 dB), the magnitude of temperature oscillations is about $10^{-4}$ °C and go undetected by human senses [1]. Working at high pressure amplitudes, the thermal interaction of sound waves with a different medium, a solid for instance can result into sufficiently large amount of heat exchange between the fluid and the solid.

## 1.2 Thermoacoustic Refrigerator (TAR)

A thermoacoustic refrigerator pumps heat from low temperature to high temperature region using energy of sound waves. Schematic of a thermoacoustic refrigerator is shown in Figure 1.1.

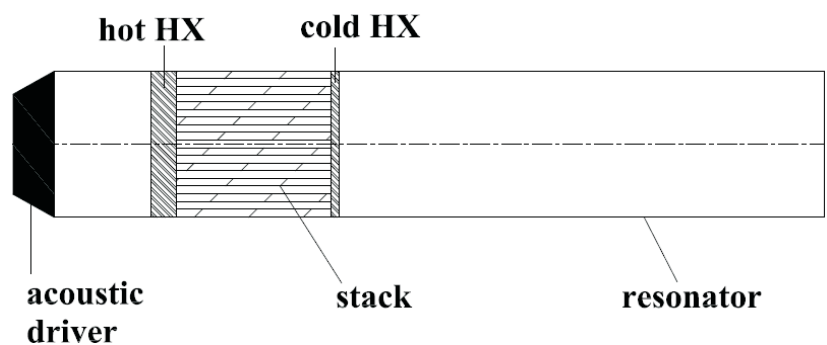

**Figure 1.1** Schematic of a thermoacoustic refrigerator.

The source of acoustic energy is called 'acoustic driver' which can be a loudspeaker. The acoustic driver emits sound waves in a long hollow tube filled with gas at high pressure. This long hollow tube is called 'resonance tube' or simply 'resonator'. The frequency of the driver and the length of the resonator are chosen so as to get a standing pressure wave in the resonator. A solid porous material like a stack of solid plates is kept in the path of sound waves in the resonator.

Due to thermoacoustic effect (explained in detail in section 2.2), heat starts to flow from one end of stack to the other. One end starts to heat up while other starts to cool down. By controlling temperature of hot side of stack (by removing heat by means of a heat exchanger), the cold end of stack can be made to cool down to lower and lower temperatures. A refrigeration load can then be applied at the cold end by means of a heat exchanger.

*'Thermoacoustics'* is a 'green' and a new technology. The working medium of TAR is an inert gas. There is no need of conventional refrigerants like CFCs that pose hazards to the environment. TAR has minimal moving parts and no valves to regulate fluid flow. Once designed efficiently, they require very less maintenance. Because of use of acoustic power, the pressure difference between which a TAR operates is very small. This means TAR can find immense application where noise or vibration can't be tolerated. Besides this, they have no close tolerances and can be fabricated from easily available materials.

## 1.3 Objectives of present work

The aim of present work is **"To design and develop a Standing Wave TAR driven by a loudspeaker, capable of cooling to a temperature near 250 K"**.

The principal objectives of the project are as stated under:

**1)** To theoretically investigate the effect of different operating parameters and working gas on TAR performance, so as to come up with a suitable TAR design.

**2)** To develop a TAR setup driven by an available electro-dynamic loudspeaker motor. This involves suitable modifications of the motor, design and fabrication of various TAR components, assembly and experimental investigations.

# Chapter 2
# LITERATURE REVIEW

## 2.1 Introduction

The two well known thermoacoustic effects- thermoacoustic pressure wave generation and thermoacoustic heat pumping are very well explained by the classical Linear Theory of Thermoacoustics [2]. The Linear Theory has also been successfully implemented in development of practical prime movers and refrigerators. This chapter begins by giving a brief insight to the above mentioned thermoacoustic effects. It then explains the Linear Thermoacoustic Theory of Thermoacoustics in detail. The chapter also presents important theoretical and experimental advances in the field of Thermoacoustic Refrigeration.

## 2.2 Thermoacoustic Oscillations

The study of thermoacoustic oscillations has a rich and interesting history. Bryon Higgins [1,3] (1777) made the first observations of thermoacoustic oscillations. A organ pipe open at both ends started to emit sound when it was heated at certain locations along its length. Sondhauss [3] (1850) made experimental investigations of heat generated sound when blowing a hot glass bulb at the end of a cold glass tube. Rijke [3] (1859) found that strong sound oscillations can be generated in an open ended hollow tube by keeping a heated wire-mesh screen at a quarter length distance from open end. The experimental configurations for study of thermoacoustic oscillations are shown in Figure 2.1.

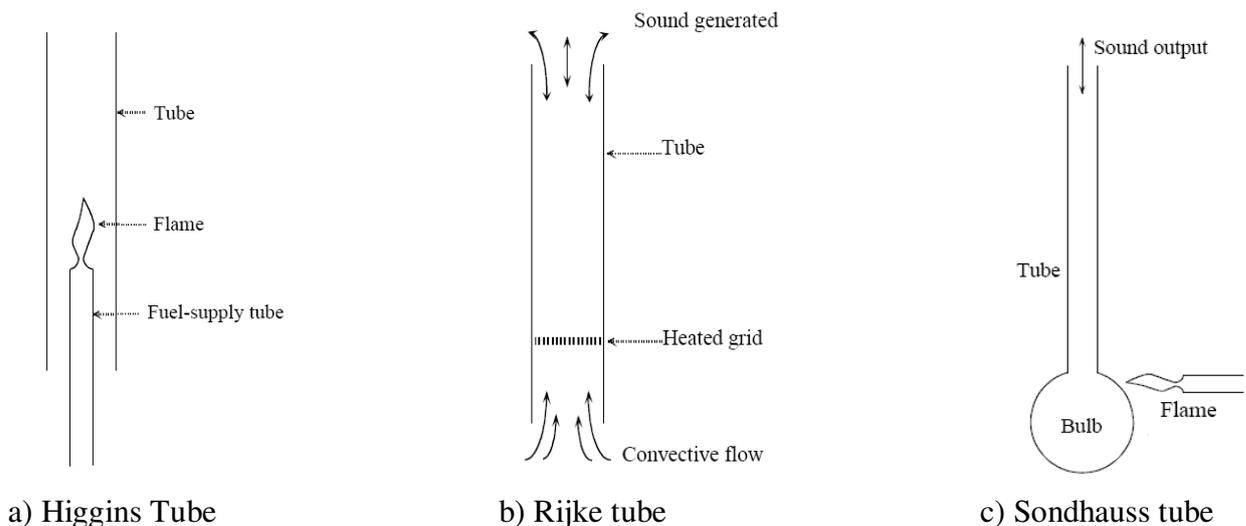

a) Higgins Tube          b) Rijke tube          c) Sondhauss tube

**Figure 2.1** Configurations for study of thermoacoustic oscillations [3]

Theoretical study of thermoacoustic oscillations began in 1868 when Kirchoff [1] calculated acoustic attenuation in a duct due to oscillatory heat transfer between solid isothermal duct wall and the gas sustaining the sound wave. The first thorough qualitative description of thermoacoustic oscillations was given by Lord Rayleigh. In his work, "The Theory of Sound" [3] (1877), he stated:

> "If heat be given to air at the moment of greatest condensation or taken from it at the moment of greatest rarefaction, the vibration is encouraged."

Yet another type of oscillations was observed by Taconis [3]. When one end of a hollow pipe was dipped in liquid nitrogen and taken out, other end being held at ambient, the pipe began to "sing". Sound was emitted continuously till the temperature of cold end became sufficiently high.

The generation of sound wave due to temperature difference across a hollow pipe can be understood from Figure 2.2. The kinetic energy of the gas near the heated area is much more than that of the gas far away in the pipe. The hot gas molecules accelerate towards the cooler end of the tube (Figure 2.2a), thereby creating an area of relative low pressure at the heated end. The cooler other gas molecules accelerate towards the hot end to fill the area of low pressure (Figure 2.2b). These molecules are then heated, and the cycle continues (Figure 2.2c). The result is a series of longitudinal air pressure oscillations. By choosing proper heat rate, heating location and length of the pipe, strong pressure oscillations can be generated.

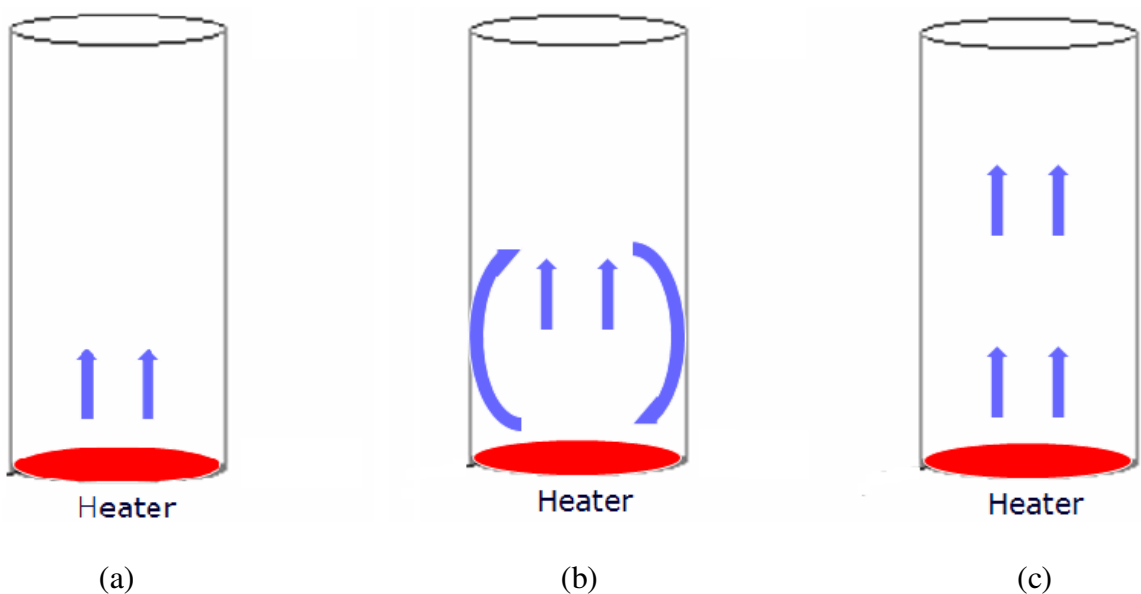

**Figure 2.2** Production of pressure oscillations in a hollow pipe by means of heat.

## 2.3 Thermoacoustic Heat Pumping

Although the production of sound or pressure pulses by application of heat has been a subject of study over last two centuries, the reverse phenomenon *i.e* achieving heat transport by means of interaction of an acoustic field with a solid medium is more recent. Heat transport process at inner wall of a hollow tube enclosing an acoustic field was demonstrated by Gifford and Longsworth [4] (1966) in their device called the "Pulse Tube". This "Basic Pulse Tube" is the precursor of widely studied modern day Pulse Tube Cryocoolers. Theoretical study of heating at the closed end of a hollow tube in which small gas oscillations were maintained, was done by Rott [2] (1984). It was based on linear theory of acoustic oscillations. The in-depth study of thermoacoustic heat pumping process was started at The Los Alamos National Laboratory (LANL) in America by The Condensed Matter and Thermal Physics Group in 1980s.

The formation of temperature gradient due to acoustic oscillations along the length of a plate can be understood from Figure 2.3 [1,4]. Consider a solid plate placed in an acoustic field with direction of particle oscillation along its length. Suppose the pressure antinode (region of maximum pressure variation) is near the left end of the plate and the pressure node (region of zero pressure variation) is near the right end of the plate. The mean temperature of the plate as well as the gas is $T_m$ and mean pressure of the gas is $p_m$. A typical gas parcel oscillates over a distance $2x_1$ about its mean position. Its pressure varies between $p_m - p_1$ and $p_m + p_1$. The activity of a typical gas parcel is shown below in Figure 2.3 (a-d).

a) 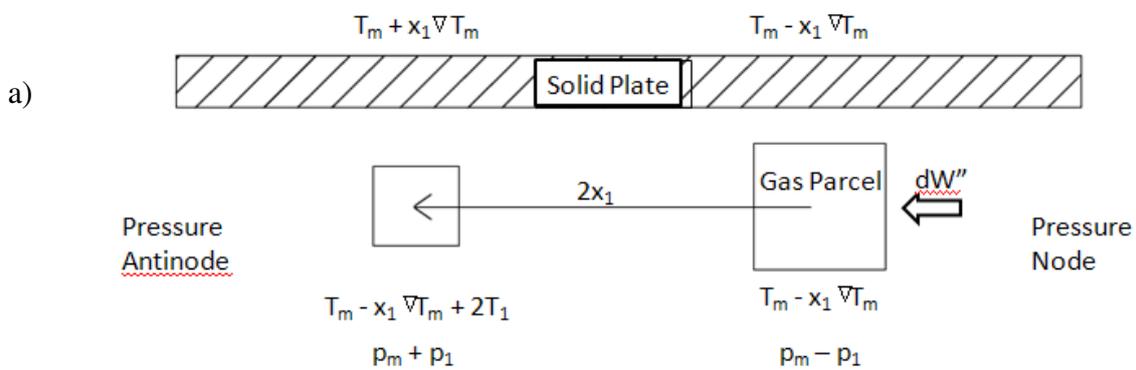

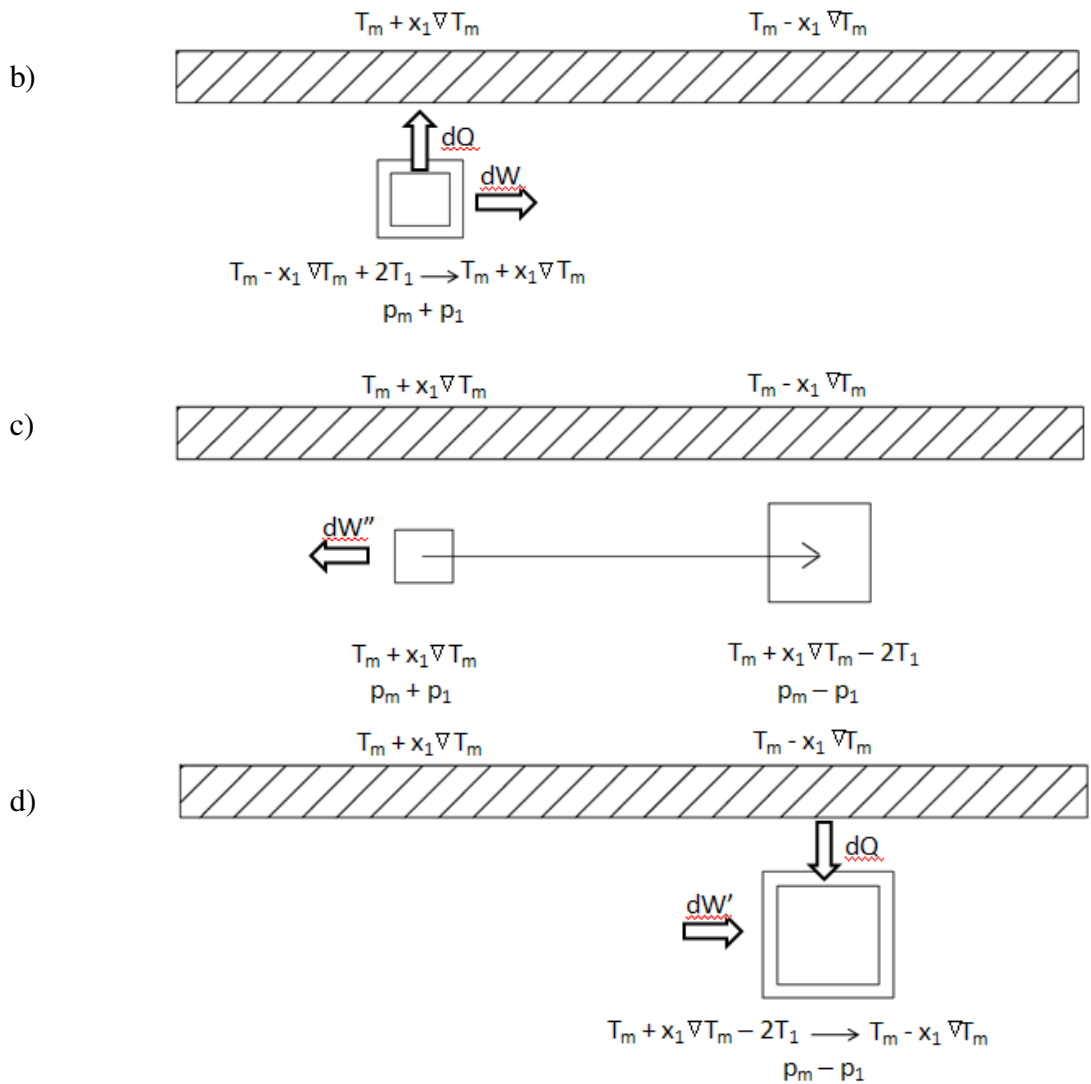

**Figure 2.3** Thermoacoustic heat pumping process.

Referring to Figure 2.3(a), the gas parcel at the right end absorbs acoustic power and moves by a distance '2x1' to the left. During this displacement it gets compressed and its temperature rises. This parcel then loses heat to the plate till its temperature equals that of the plate (Figure 2.3(b)). As a result left end of the plate becomes a little warmer. The parcel then moves again to right end where its pressure as well temperature falls (Figure 2.3(c)). The cold parcel warms up by picking heat from right end of the plate, making the right end of the plate colder (Figure 2.3(d)). Thus, in one cycle the gas transports 'dQ' amount of heat over a temperature difference of '$2x_1 \nabla T_m$', absorbing 'dW` - dW' amount of acoustic power. Eventually, the left end of stack heats up and the right end cools down.

## 2.4 The Linear Theory of Thermoacoustics

The Linear Theory of Thermoacoustics was developed by Rott [2]. It was later reviewed by Swift [1,4] who also extended it to the non-ideal cases of practical thermoacoustic prime movers and refrigerators. This section presents the theory given by Rott and Swift.

### 2.4.1 Analysis of a Single Plate

As mentioned in Chapter 1, temperature oscillations accompany pressure oscillations in an adiabatic acoustic field. Consider a solid plate kept in a fluid (a gas in general), aligned parallel to the direction of vibration of the standing wave as shown in Figure 2.4. Due to solid-fluid interaction, two phenomena occur:

    1) a time averaged heat flux near the surface along the direction of acoustic vibration

    2) absorption or generation of real acoustic power near the surface of the plate.

Suppose the length of the plate is $\Delta x$, width is $\Pi/2$ and the thickness is very small. The acoustic field vibrates along x and the local pressure and velocity of oscillation at any position x are respectively given by,

$$p_1 = p_a \sin(kx) \qquad (2.1)$$

$$u_1 = i\left(\frac{p_a}{\rho_m a}\right)\cos(kx) \qquad (2.2)$$

where 'i' represents $90^0$ phase difference between pressure and velocity oscillations due to standing wave phasing. The mean temperature of the plate as well as the fluid is $T_m$.

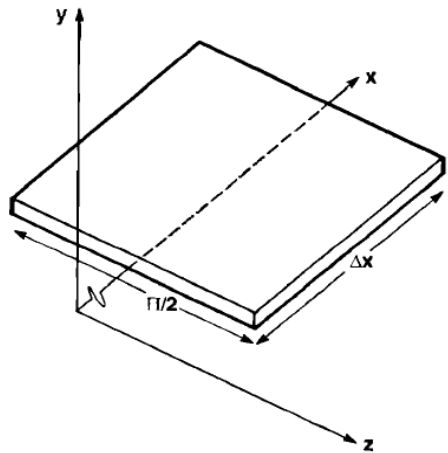

**Figure 2.4** A solid plate kept in an acoustic field. The length of plate is $\Delta x$ along x, $\Pi/2$ along z and negligible along y axis. (Reproduced from [1])

Following assumptions are made in the analysis:

1) The length of the plate is very small as compared to the wavelength of the acoustic field ($\Delta x \ll \lambda$) so that the pressure and velocity oscillations over the entire plate can be assumed to be uniform.

2) The thermal properties of gas as well as the plate do not vary with temperature.

3) The fluid is non-viscous so that the viscous boundary layer is absent and oscillatory velocity does not vary along y direction.

4) Heat capacity of the plate is very large as compared to the gas so that the gas at plate surface behaves isothermally (temperature oscillations of the gas very near to the plate are zero).

5) Heat conduction by solid as well as gas along x direction is neglected.

6) A uniform temperature $\nabla T_m$ exists along the plate in x direction.

The equation governing the energy flow through the fluid is the general equation of heat transfer given by:

$$\rho T \left( \frac{\partial s}{\partial t} + v.\nabla s \right) = \nabla.(k \nabla T) \qquad (2.3)$$

Writing eqn(2.3) in terms of oscillatory quantities $s_1$ and $T_1$ (the entropy and temperature oscillations repectively),

$$\rho_m T_m (i\omega s_1 + u_1 \frac{\partial s_m}{\partial x}) = (k \frac{\partial^2 T_1}{\partial y^2}) \qquad (2.4)$$

where $u_1$ is the oscillatory velocity in x-direction.

Substituting the oscillating entropy in terms of oscillating pressure $p_1$, and oscillating temperature $T_1$ as:

$$s_1 = \frac{c_p}{T_m} T_1 - \frac{\beta}{\rho_m} p_1 \qquad (2.5)$$

a second order differential equation in $T_1$ is obtained as given below:

$$i\omega c_p T_1 - k \frac{d^2 T_1}{dy^2} = i\omega T_m \beta p_1 - \rho_m c_p \nabla T_m u_1 \qquad (2.6)$$

Here, $\omega$ is the frequency of oscillations while $\beta$ is the thermal expansion coeffieicnt of the gas. With an isothermal boundary condition at the plate (y=0) and finite value of temperature oscillation at very large distance from plate (y=$\infty$), the expression for temperature oscillations can be found out as:

$$T_1 = \left( \frac{T_m \beta}{\rho_m c_p} p_1 - \frac{\nabla T_m}{\omega} u_1 \right)(1 - e^{-(1+i)y/\delta_k}) \qquad (2.7)$$

where
$$\delta_k = \left(\frac{2K}{\rho_m c_p \omega}\right)^{1/2} \quad (2.8)$$

is the 'thermal penetration depth' of the gas and is defined as the length of gas through which heat diffuses in time $1/\omega$. The first term in $T_1$ is due to the adiabatic compression and expansion in the fluid (as would exist if there was no plate) and the second comes into being because of the temperature gradient along the plate. At a certain value of $\nabla T_m$, the temperature oscillations vanish for all y. This value is called as the 'critical temperature gradient' and is given by,

$$\nabla T_{crit} = \frac{T_m \beta \omega p_1}{\rho_m c_p u_1} \quad (2.9)$$

The expressions for the heat flux per unit area and acoustic power absorbed/produced per unit volume or fluid surrounding the plate are given respectively by,

$$\dot{q}_2 = \frac{1}{2}\rho_m c_p \operatorname{Im}[T_1] u_1 \quad (2.10)$$

$$\dot{w}_2 = -\frac{1}{2}\omega \beta p_1 \operatorname{Im}[T_1] \quad (2.11)$$

Both the quantities depend on imaginary part of $T_1$. As can be seen from Figure 2.5, the imaginary part of $T_1$ starts from zero at y=0, becomes maximum at y~$\delta_k$ and again vanishes for y>> $\delta_k$. That is to say, the heat flux and the acoustic power absorbed/generated are predominant in the region y~$\delta_k$ from the plate surface. Both vanish at distances very close and very far from the plate.

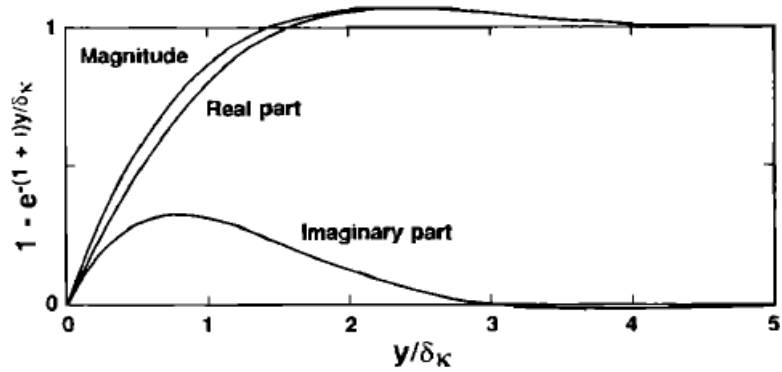

**Figure 2.5** The real and imaginary parts of $T_1(y)$. (Reproduced from [1])

The total heat flux and the acoustic power absorbed/generated can be found by integrating eqns. (2.10) and (2.11) in whole of the region surrounding the plate. They are given by,

$$\dot{Q}_2 = -\frac{1}{4}\Pi \delta_k T_m \beta p_1 u_1 (\Gamma - 1) \quad (2.12)$$

$$\dot{W}_2 = \frac{1}{4}\Pi\delta_k\Delta x \frac{T_m\beta^2\omega}{\rho_m c_p}p_1^2(\Gamma-1) \tag{2.13}$$

where $\Gamma$ is the ratio of actual temperature gradient to the critical temperature gradient. An important point is to be noted here. When $\Gamma<1$ (a small temperature gradient along the plate which is less than the critical temperature gradient), the expression of $W_2$ has a negative sign indicating that acoustic power is being absorbed near the plate. In this case, $Q_2$ is positive i.e heat is being transported from pressure node to pressure antinode (heat pumping). On the other hand, when there is a large temperature gradient along the plate ($\Gamma>1$), heat flows from pressure antinode to pressure node and acoustic power is produced. In a very special case when $\Gamma=1$, no power is absorbed or produced and the heat flux is zero. Hence, three modes of operation can be classified:

1) $\Gamma<1$ : Heat pump or refrigerator (acoustic power getting absorbed)
2) $\Gamma>1$ : Prime mover (acoustic power getting generated)
3) $\Gamma=1$ : No practical significance (acoustic power is zero)

## 2.4.2 Analysis of a Stack of Parallel Plates

In this section, the analysis of a single plate is extended to a stack of parallel plates as shown in Figure 2.6. The fluid is assumed to have arbitrary viscosity with Prandtl number σ, and the fluid and solid plates respectively have thermal conductivities K and $K_s$. Thus, this analysis brings the Linear Theory of Thermoacoustics closer to real refrigerators and prime movers.

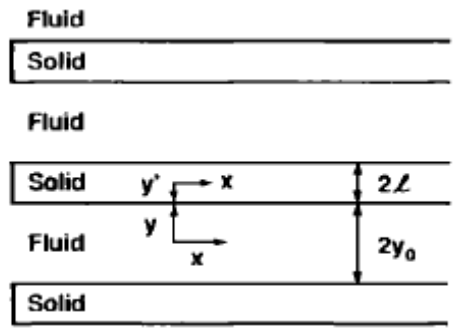

**Figure 2.6** A stack of parallel plates. Each plate has a thickness 2l and spacing between plates is $2y_0$. (Reproduced from [1])

The solid plates have a thickness 2l and spacing between plates is $2y_0$. The stack is oriented in the direction of acoustic oscillations x. The equations governing the analysis are the continuity equation:

$$\frac{\partial\rho}{\partial t}+\nabla\cdot(\rho v)=0 \tag{2.14}$$

the momentum equation:

$$\rho\left(\frac{\partial v}{\partial t} + (v \cdot \nabla)v\right) = -\nabla p + \mu \nabla^2 v \qquad (2.15)$$

and the energy equations for the fluid and the plate:

$$\rho T(\frac{\partial s}{\partial t} + v \cdot \nabla s) = \nabla \cdot (k\nabla T) \qquad (2.16)$$

$$\rho_s c_s \frac{\partial T_s}{\partial t} = k_s \nabla^2 T_s \qquad (2.17)$$

The time dependent variables appearing in the above equations can be written as:

$$p = p_m + p_1(x)e^{i\omega t} \qquad (2.18)$$

$$\rho = \rho_m(x) + \rho_1(x,y)e^{i\omega t} \qquad (2.19)$$

$$v = \hat{x}u_1(x,y)e^{i\omega t} + \hat{y}v_1(x,y)e^{i\omega t} \qquad (2.20)$$

$$T = T_m(x) + T_1(x,y)e^{i\omega t} \qquad (2.21)$$

$$T_s = T_m(x) + T_{s1}(x,y)e^{i\omega t} \qquad (2.22)$$

$$s = s_m(x) + s_1(x,y)e^{i\omega t} \qquad (2.23)$$

Substituting these expressions into the governing equations and integrating the resulting expressions in the stack region yields the Thermoacoustic wave equation:

$$\left(1 + \frac{(\gamma-1)f_k}{1+\varepsilon_s}\right)p_1 + \frac{\rho_m a^2}{\omega^2}\frac{d}{dx}\left(\frac{1-f_v}{\rho_m}\frac{dp_1}{dx}\right) - \beta\frac{a^2}{\omega^2}\frac{f_k - f_v}{(1-\sigma)(1+\varepsilon_s)}\frac{dT_m}{dx}\frac{dp_1}{dx} = 0 \qquad (2.24)$$

and the equation of energy flux through stack cross section:

$$\dot{H}_2 = \frac{\Pi y_0}{2\omega\rho_m}\text{Im}\left[\frac{d\tilde{p}_1}{dx}p_1\left(1 - \tilde{f}_v - \frac{T_m\beta(f_k - \tilde{f}_v)}{(1+\varepsilon_s)(1+\sigma)}\right)\right] + \frac{\Pi y_0 c_p}{2\omega^3\rho_m(1-\sigma)}\frac{dT_m}{dx}\frac{dp_1}{dx}\frac{d\tilde{p}_1}{dx}$$

$$\times \text{Im}\left[\tilde{f}_v + \frac{(f_k - \tilde{f}_v)(1+\varepsilon_s f_v/f_k)}{(1+\varepsilon_s)(1+\sigma)}\right] - \Pi(y_0 k + lk_s)\frac{dT_m}{dx} \qquad (2.25)$$

where $f_k$ and $f_v$ are the Rott's functions for temperature and viscosity given by:

$$f_k = \frac{\tanh[(1+i)y_0/\delta_k]}{(1+i)y_0/\delta_k} \qquad (2.26)$$

$$f_v = \frac{\tanh[(1+i)y_0/\delta_v]}{(1+i)y_0/\delta_v} \qquad (2.27)$$

and $\varepsilon_s$ is the heat capacity ratio of the fluid-solid system,

$$\varepsilon_s = \frac{\rho_m c_p \delta_k \tanh[(1+i) y_0 / \delta_k]}{\rho_s c_s \delta_s \tanh[(1+i) y_0 / \delta_s]} \tag{2.28}$$

Due to viscosity of the gas, the viscous penetration depth comes into picture. It is the distance from plate surface in which the viscous effects are predominant. The viscous penetration depth is given by,

$$\delta_v = \sqrt{\frac{2\mu}{\rho_m \omega}} \tag{2.29}$$

## 2.4.3 The Boundary Layer Approximation

The boundary layer approximation [1,4] states that the half plate spacing is large as compared to the thermal penetration depth of the gas ($y_0 \gg \delta_k$) and the half plate thickness is large as compared to the thermal penetration depth of the solid plate ($l \gg \delta_s$). The use of this approximation is to set the hyperbolic tangents appearing in eqns. (2.26-2.28) equal to one which simplifies the thermoacoustic wave and energy flux equation to a great extent. These equations with boundary layer approximation are given by:

$$p_1 + \frac{\rho_m a^2}{w_2} \frac{d}{dx}\left(\frac{1-f_v}{\rho_m} \frac{dp_1}{dx}\right) = \frac{(\gamma-1)\delta_k p_1}{(1+i)(1+\varepsilon_s) y_0}\left(\frac{\Gamma}{(1+\sqrt{\sigma})(1-f_v)} - 1\right) \tag{2.30}$$

$$\dot{H}_2 = -\frac{1}{4}\Pi \delta_k \frac{T_m \beta p_1 u_1}{(1+\varepsilon_s)(1+\sigma)(1-\delta_v/y_0 + \delta_v^2/2y_0^2)}$$
$$\times [\Gamma \frac{1+\sqrt{\sigma}+\sigma+\sigma\varepsilon_s}{1+\sqrt{\sigma}} - (1+\sqrt{\sigma} - \delta_v/y_0)] - \Pi(y_0 k + l k_s)\frac{dT_m}{dx} \tag{2.31}$$

The first term in eqn. (2.30) is the hydrodynamic flow of heat due to the thermoacoustic effect while the second term accounts for the heat flow along the stack due to conduction in gas and solid. The net acoustic power absorbed/generated in the stack with the boundary layer approximation is given by:

$$\dot{W}_2 = \frac{1}{4}\Pi \delta_k \Delta x \frac{(\gamma-1)\omega p_1^2}{\rho_m a^2 (1+\varepsilon_s)} \times \left(\frac{\Gamma}{(1+\sqrt{\sigma})(1-\delta_v/y_0 + \delta_v^2/2y_0^2)} - 1\right) \tag{2.32}$$
$$-\frac{1}{4}\Pi \delta_v \Delta x \frac{\omega \rho_m u_1^2}{(1-\delta_v/y_0 + \delta_v^2/2y_0^2)}$$

In eqn.(2.31), the first term represents the acoustic power absorbed/generated in the stack while the second term accounts for the dissipation of acoustic power in the viscous layer which is converted to heat (a loss).

## 2.4.4 Arbitrary Stack Geometry

A general formulation of thermoacoustics for stacks having arbitrarily shaped pore cross section was given by Arnott *et.al.* [5]. Expressions for oscillatory temperature, pressure and velocity were formulated for a stack with arbitrary shaped pores. Using these expressions, the heat and work flows in geometries like parallel plate, circular pores, hexagonal pores, equilateral triangular pores and rectangular pores were developed and compared. It was concluded that the parallel plate stack gave optimum heat and work flows. The stack geometries studied by Arnott *et.al* are shown in Figure 2.7

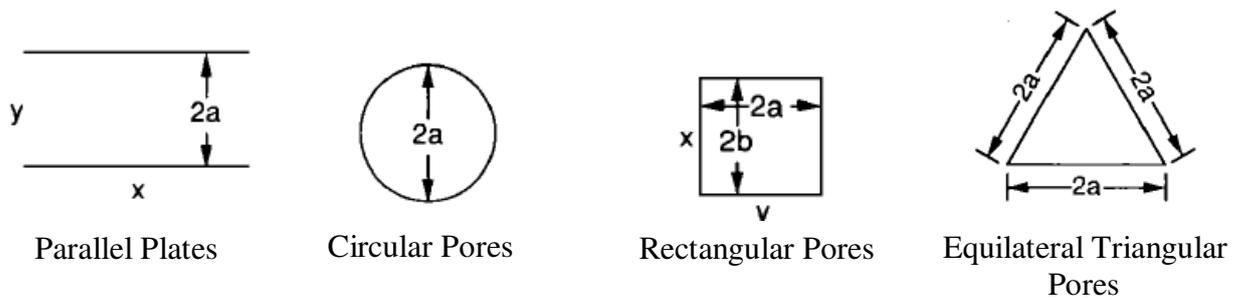

**Figure 2.7** Various pore geometries studied by Arnott *et.al.*(Reproduced from [4])

A new stack geometry, 'pin array' was analyzed by Keolian *et.al.*[6]. Analytical expressions for the oscillatory temperature and velocity for the pin array geometry were derived by using Arnott's general formulation. It was shown for given set of parameters, the performance of pin array stack was better than other geometries. The unit cell of pin array geometry and the Rott's function for various geometries are shown in Figure 2.8 and Figure 2.9 respectively.

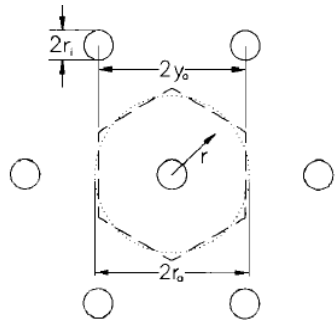

**Figure 2.8** The hexagonal unit cell of a pin array stack. (Reproduced from [5])

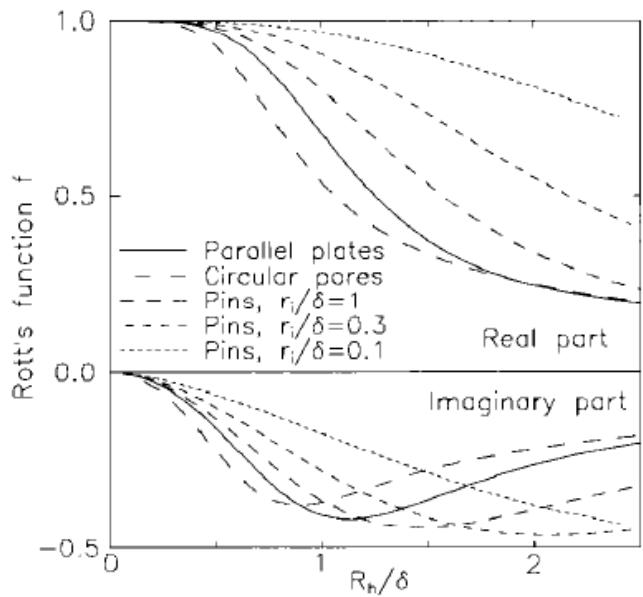

**Figure 2.9** Real and imaginary parts of Rott's functions for parallel plates, circular pores and pin array stack. (Reproduced from [5])

## 2.5 Thermoacoustic Refrigerators (TARs)

Swift's review paper [1] led to development of many practical thermoacoustic refrigerators. Some of important theoretical and experimental findings are described in this section.

### 2.5.1 Theoretical Models

A design algorithm for TAR was given by Herman *et.al* [7]. Various operating and design parameters were indentified by the authors as shown in Table 2.1. To reduce the complexity of theoretical expressions due to a large number of variables, normalization technique was implemented.

**Table 2.1** Operating, design parameters and material properties

| Operating Parameters | | Design Parameters | | | | Material Properties | | | |
|---|---|---|---|---|---|---|---|---|---|
| | | Design requirements | | Stack geometry | | Working gas | | Stack | |
| $p_m$ | Mean Pressure | $T_c$ | Cold end Temperature | $L_s$ | Stack length | $\mu$ | Dynamic Viscosity | $\rho_s$ | Density |
| $f$ | Operating Frequency | $T_h$ | Hot end Temperature | $x_s$ | Stack center position | $k$ | Thermal Conductivity | $k_s$ | Thermal conductivity |
| $p_1$ | Pressure Amplitude | $Q_c$ | Cooling Load | $l$ | Half Plate Thickness | $\gamma$ | Cp/Cv ratio | $C_s$ | Specific Heat Capacity |
| $T_m$ | Mean Temperature | $W$ | Input Acoustic Power | $y_0$ | Half Plate spacing | $a$ | Sound Speed | | |
| | | | | $A$ | Cross section | | | | |

The simplified expressions for cooling power and acoustic power in terms of normalized parameters are respectively given by,

$$Q_{cn} = -\frac{\delta_{kn} D^2 \sin(2x_{ns})}{8\gamma(1+\sigma)\Lambda} \times \left( \frac{\Delta T_{mn} \tan(x_{ns})}{(\gamma-1)BL_{sn}} \frac{1+\sqrt{\sigma}+\sigma}{1+\sqrt{\sigma}} - (1+\sqrt{\sigma}-\sqrt{\sigma}\delta_{kn}) \right) \quad (2.33)$$

$$W_n = \frac{\delta_{kn} L_{sn} D^2}{4\gamma}(\gamma-1)B\cos^2(x_{ns}) \times \left( \frac{\Delta T_{mn} tan(x_{ns})}{BL_{sn}(\gamma-1)(1+\sqrt{\sigma})\Lambda} - 1 \right) \quad (2.34)$$

$$-\frac{\delta_{kn} L_{sn} D^2}{4\gamma} \frac{\sqrt{\sigma}\sin 2(n_{xs})}{B\Lambda}$$

where
$$\Lambda = 1 - \sqrt{\sigma}\delta_{kn} + 0.5\sigma\delta_{kn}^2 \quad (2.35)$$

The design algorithm is as follows:

**Step 1:** Choose the operating parameters such as mean pressure (with the constraint of material strength), mean temperature, frequency, and the pressure amplitude. The pressure amplitude should be low enough so as not to induce turbulence. This is ensured by keeping the acoustic Reynold's number less than 500 [1]. The acoustic Reynold's number is given by,

$$\text{Re}_a = \frac{u_1 \delta_v \rho}{\mu} \quad (2.36)$$

Choose the working gas and stack material. From the calculated values of thermal penetration depth, find plate spacing of the stack. The optimized plate spacing is equal to twice the thermal penetration depth [1]. Assuming porosity, find plate thickness.

**Step 2:** Define stack COP as COP=$Q_{cn}/W_n$. The stack COP becomes a function of only two variables *viz.* $L_{sn}$ and $x_{sn}$. For different values of $x_{ns}$, the values of COP are plotted with $L_{sn}$ and optimal COP in each case is found out.

**Step 3:** Choose an optimum COP from the set and corresponding $L_{sn}$ and $x_{sn}$. From the values of required cooling power Q and expression of normalized cooling power, calculate stack cross section area.

**Step 4:** Calculate the resonator length so as to obtain a standing wave phasing between oscillatory pressure and velocity. The length-frequency relation is given by,

$$L = n\frac{a}{4f} \qquad n = 1, 3, 5, \ldots \quad (2.37)$$

**Step 5:** Determine the lengths of cold and hot side heat exchangers using the values of displacement amplitudes and the cold and hot exchanger locations respectively. The optimal length equals twice the displacement amplitude [1]. The porosity of heat exchangers should be equal to that of stack to prevent any discontinuity in gas flow passage.

**Step 6:** Apply 1$^{st}$ Law to resonator-stack-HXs system and choose a suitable loudspeaker/acoustic driver which can pump in the required acoustic power.

A computational model of thermoacoustic refrigerators for performance prediction at steady state was given by Jebali *et.al.* [8]. The components of refrigerator were represented in terms of acoustic network elements *viz.* compliance, inertance, viscous resistance, thermal relaxation conductance and attenuation factors. The one dimensional cross section averaged equations were discretized using the network analogy. The frequency response of the model indicated maximum cooling power near the resonance frequency. A similar study was also made by Qiu *et.al.* [9]

Worlikar *et.al* [10,11] developed a low Mach number numerical model for simulating the flow inside the thermoacoustic refrigerator stack. The dependence of energy loses due to stack configurations and operating conditions was examined.

A simulation program DELTAEC (Design Environment for Low Amplitude Thermoacoustic Energy Conversion) was developed by Ward and Swift [12]. Based on DELTAEC, a design optimization program was developed by Paek *et.al.* [13]. It was shown that COPR of thermoacoustic refrigerators is maximum at about 80 K temperature lift.

An analytical model to study transient temperature profile inside thermoacoustic refrigerator stack was given by Lotton *et.al* [14]. The contribution of various heat transfer processes like thermoacoustic heat transport, longitudinal conduction, radial heat leakage, heating due to vorticity etc during the transient regime was studied. The analytical results were obtained by fitting in empirical heat transfer coefficients. It was verified that at steady state, the thermoacoustic heat flux is balanced by the returning longitudinal conduction.

A model describing the heat transfer between elements of thermoacoustic refrigerators- stack and heat exchangers was given by Brewster *et.al* [15]. Forced heat convection due to acoustic oscillations was identified as the main heat transfer mechanism. For low amplitudes of oscillations, the magnitude of heat transfer coefficient was proportional to the local amplitude of oscillatory velocity.

## 2.5.2 Experimental Work

The first fully functional thermoacoustic refrigerator was designed and built by Tom Hofler [16,17]. The loudspeaker driven refrigerator employed Kapton as the material for stack and Helium as the working gas. Lowest temperature ratio achieved by Hofler was 0.66 while the optimum COP was 12 % of Carnot COP. The operating parameters and stack dimensions of Hofler's refrigerator are given in Table 2.2.

Table 2.2 Hofler's operating parameters and stack dimensions

| Operating parameters | | Stack Dimensions | |
|---|---|---|---|
| Mean Pressure | 10 bar | Plate Thickness | 0.08 mm |
| Frequency | 500 Hz | Plate Spacing | 0.38 mm |
| Drive Ratio | 3 % | Length | 0.08 m |
| Mean Temperature | 255 K | Cross Section Area | 0.0012 m$^2$ |
| Hot end temperature | 300 K | Centre Location | 0.09 m |

Based on the design algorithm given by Herman *et.al.*, Tijani *et.al.* designed [18] and constructed [19] a loudspeaker driven thermoacoustic refrigerator. The effect of blockage ratio on performance of refrigerator was studied experimentally by varying the plate spacing [20]. It was observed that maximum heat flux occurs when plate spacing was twice the thermal penetration depth. By using binary mixtures of inert gases [21], the effect of Prandtl number was studied. A 30-70 mixture of Xe-He with a Prandtl Number of 0.2 gave optimum results. A technique to optimize the loudspeaker to drive a given refrigerator was given by the authors [22]. The lowest temperature reported was -65$^o$C. The operating parameters and stack dimensions are given below in Table 2.3.

Table 2.3 Tijani's operating parameters and stack dimensions

| Operating parameters | | Stack Dimensions | |
|---|---|---|---|
| Mean Pressure | 10 bar | Plate Thickness | 0.1 mm |
| Frequency | 400 Hz | Plate Spacing | 0.3 mm |
| Drive Ratio | 2 % | Length | 0.085 m |
| Mean Temperature | 250 K | Cross Section Area | 0.00118 m$^2$ |
| Hot end temperature | 283 K | Centre Location | 0.08 m |

Adeff *et.al.* [23] used reticulated vitreous carbon (RVC) as the stack material. The performance of RVC stack was found comparable to plastic roll stack and the lowest temperature ratio obtained was 0.82.

An experimental research to study the effect of operating parameters was carried out by Nsofor *et.al.* [24]. The maximum cooling load was obtained at the resonance frequency. It was also shown that an optimal mean pressure existed for maximal cooling load.

Scalability of thermoacoustic refrigerators was studied analytically and experimentally by Li *et.al.* [25]. It was found that scaling down of thermoacoustic refrigerators was limited by heat conduction.

A flow through thermoacoustic refrigerator was developed by Reid *et.al.* [26]. A steady mass flow was inlet to refrigerator at one of the pressure nodes and was outlet from another. A 20 % increase in COP was observed as compared to closed refrigerators.

Besides these, several other refrigerators were developed by various research groups. Details of some notable refrigerators are given in Table 2.4.

**Table 2.4** Performance of thermoacoustic refrigerators

| Name | $\Delta T$ ($^o$C) | $T_c$ ($^o$C) | $Q_c$ (W) | COPR |
|---|---|---|---|---|
| Prototype at Purdue University [13] | 8.9 | 15.6 | 40 | 0.033 |
| Frankenfridge [27] | 10.8 | 17.2 | 26 | 0.067 |
| Triton [13] | 18 | 10.4 | 2161 | 0.04 |
| Ben & Jerry cooler [13] | 58.5 | -24.6 | 119 | 0.22 |

## 2.6 Summary

It can be summarized from the literature that the theory of standing wave thermoacoustics is well established. Several theoretical models as well as simulation softwares to forecast the performance of a TAR at steady state are also available. However, there are only a few notable instances [3,13,16,27] where a fully functional standing wave TAR is designed, constructed and experimentally investigated. Similarly, there is only one instance available [14], where the transient state temperature profiles inside the stack of a TAR are reported.

In view of this, it is decided to design and develop a standing wave thermoacoustic refrigerator capable of cooling to temperatures near 250 K. Due to time limited design data on acoustic drivers, a commercially available electro-dynamic motor will be used. All other components of the refrigerator would be designed to adapt to the available motor. The effect of operating parameters on TAR performance at steady as well as transient state would be investigated theoretically and experimentally.

# Chapter 3

# THEORETICAL ANALYSIS OF A STANDING WAVE TAR

## 3.1 Introduction

A standing wave TAR can be divided into two major sub-systems – the heat pumping assembly and the driver assembly. The heat pumping system consists of the stack, the two heat exchangers and the resonator. The driver assembly consists of a magnet-voice coil based electro-dynamic motor and a pusher cone suitably constructed for thermoacoustic refrigeration.

This chapter is divided into three parts. The first part describes a theoretical model of a loudspeaker driven straight gas column. This can be thought of a TAR assembly void of the stack and the heat exchangers. The resonance frequency and dynamic pressure generated in the gas column can be predicted using this analysis. Due to observed non-linear effects in straight resonators, it is concluded that a resonator with non-uniform cross section be used. The second part of this chapter describes the design of this new resonator using an available software, DeltaEC. Finally, in the third part, an iterative method is developed to predict the thermodynamic performance of a TAR configuration.

## 3.2 Loudspeaker Driven Gas Columns

In order to predict the thermodynamic performance of a TAR, it is first essential to determine the dynamic pressure that the available acoustic driver can generate. Other than the driver parameters which are introduced in following sub-section, the dynamic pressure depends on the charging pressure, the working gas and the operating frequency of the system. The theoretical prediction of the dynamic pressure is done using the impedance transfer technique. This technique enables one to calculate the effective acoustic impedance of the load at the driver piston, and its resonance frequency. The electrical network model of a moving coil loudspeaker is coupled with this acoustic impedance model so that the dynamic pressure can be directly computed as a function of input electrical voltage to the system.

## 3.2.1 The Impedance Transfer Technique

Consider a straight hollow channel of length 'L' and uniform cross section area 'A', filled with a gas at pressure '$p_m$' and temperature '$T_m$'. The speed of sound in the gas is 'a'. A source of acoustic wave (acoustic driver) is attached to one end of the channel at x=0. When an acoustic wave with an angular frequency 'ω' propagates through this channel, the acoustic impedance at any location 'x' in the channel is given by:

$$Z_{ac}(x) = \frac{p_1(x)}{Au_1(x)} \tag{3.1}$$

where, $p_1(x)$ and $u_1(x)$ are the oscillatory pressure and velocity at location 'x'. The transfer function giving the acoustic impedance at location 'x' in terms of acoustic impedance at any other location 'x`' is given by [3]:

$$Z_{ac}(x) = \frac{Z_{ac}(x')\cos k(x'-x) + jZ_c \sin k(x'-x)}{j\frac{Z_{ac}(x')}{Z_c}\sin k(x'-x) + \cos k(x'-x)} \tag{3.2}$$

where,

$$Z_c = \frac{\rho_m \omega}{Ak(1-f_v)} \tag{3.3}$$

Here, 'k' is the wave number and '$f_v$' is the complex Rott's viscosity function [1] denoting the loss of acoustic power at the walls of the channel. A rigidly sealed end at x=L would result in a location of infinite acoustic impedance. In this case, the acoustic impedance at the driver piston would be:

$$Z_{ac} = -j\frac{\rho_m \omega}{Ak(1-f_v)}\cot kL \tag{3.4}$$

This complex acoustic load resonates when its imaginary part is zero. The relation between the channel length, the frequency and the sound speed for fundamental resonance mode is:

$$L = \frac{\pi a}{\omega} \tag{3.5}$$

As can be seen from eqn(3.4), the acoustic impedance of a given geometry depends on the density of the working medium. Thus, it can be varied by changing the mean pressure, mean temperature or the medium itself. However in practice, it is not feasible to vary the mean temperature of the working medium.

## 3.2.2 Electrical Network Model of a Moving Coil Loudspeaker

A simple moving coil loudspeaker consists of an electrical conductor coil suspended in a radial magnetic field. When excited by an alternating current, the coil reciprocates perpendicular to the plane containing the magnetic field and current, producing a Lorentz's force. A diaphragm or a piston attached to the reciprocating coil causes periodic compressions and rarefactions in the surrounding medium, thereby producing an acoustic wave. A representation of moving coil loudspeaker in the electrical network form is shown in Figure 3.1 below:

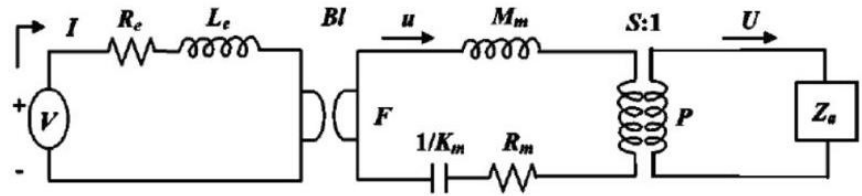

**Figure 3.1** Electrical circuit representation of a loudspeaker with an acoustic load.

Traversing from right to left in Figure 3.1, the first element is the acoustic impedance of the load at the driver piston. The centre block represents the mechanical part consisting of the moving mass $M_m$, stiffness $K_m$ and mechanical resistance $R_m$ of the loudspeaker. The leftmost block is the electrical part, comprising of the electrical resistance $R_e$ and inductance $L_e$ of the coil. The transduction coefficient 'Bl' converts the electrical energy to mechanical energy, while the piston surface area 'S' converts mechanical energy into acoustic energy. The electro-acoustic efficiency of a loudspeaker can be maximized by making the mechanical part and the acoustic part resonate at a same driving frequency [28]. The total electrical impedance of the loudspeaker-acoustic load circuit between the two terminals of ac power source is given by:

$$Z_e = R_e + j\omega L_e + \frac{(Bl)^2}{R_m + j\left(\omega M_m - \frac{k_m}{\omega}\right) + S^2 Z_{ac}} \quad (3.6)$$

## 3.2.3 Dynamic Pressure in a Loudspeaker Driven gas Column

Using the standard transduction relations, the expression for the pressure amplitude generated at the piston surface can be written as:

$$p_1 = V \frac{S(Bl)(Z_{ac})}{Z_e \left[ R_m + j\left(\omega M_m - \frac{k_m}{\omega}\right) + S^2 Z_{ac} \right]} \tag{3.7}$$

where 'V' is the voltage input to the system.

The theoretical results of this model are presented and compared with experimental results in a later chapter.

## 3.3 DeltaEC Model

In the course of experiments with loudspeaker driven gas columns, periodic shocks can be observed. This non-linear effect hamper the magnitude and distort the sinusoidal nature of dynamic pressure. Saenger and Hudson [29] have also observed such shocks which are bound to exist in straight gas columns with uniform cross section, when excited at high amplitudes and operated very near the resonance frequency. Though the shocks are minimal for low excitation levels, the dynamic pressure generated in that case is too low to drive a thermoacoustic refrigerator. It was also confirmed from [3,16,29] that such shocks can be prevented by use of resonators which are not straight, but of non-uniform cross section.

One such geometry of resonator is shown below in Figure 3.2. It consists of a larger diameter tube of diameter $D_1$ and length $L_1$, a conical reducer of length $L_2$, a smaller diameter tube of diameter $D_3$ and length $L_3$, and a conical buffer volume of length $L_4$ in specified order. The larger diameter of the conical buffer is $D_4$. The acoustic driver is located at the beginning of the larger diameter tube at x=0. A similar geometry was also used by Tijani et.al [3] in their standing wave TAR.

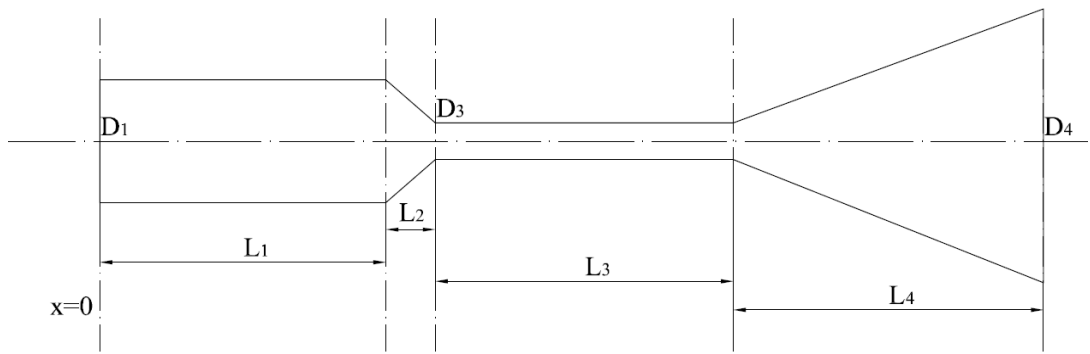

**Figure 3.2** Schematic of a resonator with a non-uniform cross section

The geometry shown above is modeled in DeltaEC. For a set of user specified operating parameters and geometry, DeltaEC integrates the thermoacoustic wave equation (eqn. 2.24) so that the specified boundary conditions are satisfied. In the present DeltaEC model, the operating parameters *viz.* the mean pressure and temperature, the dynamic pressure at x=0, the operating frequency and the working gas are specified. All the geometric parameters shown in the Figure 3.2, except $L_3$, are also assumed. The rigid termination at the extreme end of the buffer volume is a region of zero oscillatory velocity, hence a location of infinite acoustic impedance. In order to enforce resonance at the chosen operating frequency, the oscillatory pressure and velocity at x=0 are kept in phase. With these input parameters, DeltaEC solves for the value of $L_3$ and the oscillatory velocity at x=0.

An equally viable alternative is of choosing $L_3$ arbitrarily and solving for $L_4$ with the same boundary conditions. In a similar way, an iterative method may be formulated to geometrically optimize the resonator for a given set of operating parameters and boundary conditions. The geometrical parameters of this resonator model are given in the last section of this chapter.

## 3.4 Transient State Model of a Standing Wave TAR

A one dimensional model is developed to predict the transient state response of a Standing Wave TAR. The total power flux equation is used to describe the transient regime. In case of standing wave refrigerators, the contribution of travelling wave component (the acoustic power flow) to the total power flow is very negligible as compared to the standing wave component. Hence, the total power flow can be used to approximate the enthalpy flow through the stack region. The method also predicts the steady state cold temperature that a given TAR configuration can attain.

### 3.4.1 Geometry

The geometry considered here is a typical Hofler style standing wave TAR. As shown in Figure 3.3, it consists of a resonator of length L, with a source of acoustic waves (acoustic driver) kept at x=0. The end of resonator (x=L) is rigidly closed. A parallel plate stack of length $L_s$ is kept inside the resonator at $x=x_s$ from the driver end. Hot and cold heat exchangers of lengths $L_h$ and $L_c$ respectively are kept to left and right of the stack.

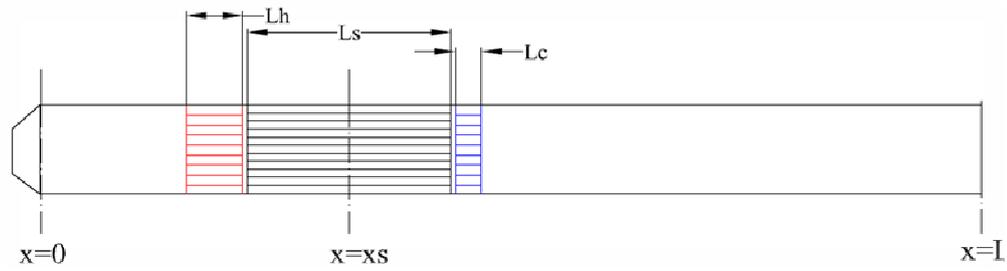

**Figure 3.3** Schematic of the geometry for Transient State Analysis

### 3.4.2 Assumptions

Following assumptions are made while carrying out the transient state analysis of the TAR :-

**1.** The model developed is One Dimensional-

**a)** Frequency of acoustic wave is small enough to ensure a plane wave-front inside the resonator. The condition for plane wave propagation thorough a hollow pipe with diameter D is [3]:

$$f < \frac{1.841a}{\pi D} \qquad (3.8)$$

**b)** Heat flow in the stack and the resonator are longitudinal, lateral heat leaks into the refrigerator are neglected

**2.** The model remains in framework of the Linear Theory with the Boundary Layer Approximation [1].

**3.** Parallel plate geometry is considered for stack as well as for the heat exchangers. Their porosities are assumed to be same and large enough not to disturb the standing wave phasing.

**4.** Complete heat exchange limit [Brewster et.al] is taken while calculating the heat transfer coefficients between the stack and the heat exchangers.

**5.** The working gas is ideal.

### 3.4.3 Governing Equations

The acoustic field in the resonator has standing wave phasing between the dynamic pressure and oscillatory velocity, with a pressure antinode at the driver end. Assuming that the presence of stack does not alter the standing wave phasing, the spacial variation for dynamic pressure and oscillatory velocity are given by:

$$p_1(x) = p_a \cos(kx) \tag{3.9}$$

$$u_1(x) = \frac{p_a}{\rho_m a} \sin(kx) \tag{3.10}$$

Due to porosity of stack, the oscillatory velocity inside the stack gets modified to:

$$u_1(x) = \frac{p_a}{\rho_m a}\left(\frac{y_0 + l}{y_0}\right)\sin(kx) \tag{3.11}$$

According to Rott's acoustic approximation [2], the time averaged heat flux at a certain cross section inside the stack is given by:

$$\dot{H}_2 = -\frac{1}{4}\Pi\delta_k \frac{T_m \beta p_1 u_1}{(1+\varepsilon_s)(1+\sigma)(1-\delta_v/y_0 + \delta_v^2/2y_0^2)} \tag{3.12}$$

$$\times[\Gamma\frac{1+\sqrt{\sigma}+\sigma+\sigma\varepsilon_s}{1+\sqrt{\sigma}} - (1+\sqrt{\sigma}-\delta_v/y_0)] - \Pi(y_0 K + l K_s)\frac{dT_m}{dx}$$

where β is the thermal expansion coefficient of working gas and Γ is the ratio of actual temperature gradient to the critical temperature gradient. The first term in RHS of eqn(3.12) is the heat pumping due to thermoacoustic effect, while the second term denotes the heat flux returning back through the stack due to conduction by solid as well as gas. For an ideal gas, the thermal expansion coefficient is the inverse of its absolute temperature. Hence, a simplified expression for heat flux can be written down as:

$$H_2(x) = A(x)\frac{dT(x)}{dx} + B(x) \tag{3.13}$$

where, A(x) and B(x) are functions of the axial coordinate in the stack.

The heat transfer coefficients between the heat exchangers and ends of stack are calculated using 'complete heat exchange limit' [15]. The heat transfer coefficients at the cold and hot heat exchangers are given by:

$$h_{HHX} = \frac{2}{\pi} A_r B_r c_p \rho_m u_{1,HHX} \tag{3.14}$$

$$h_{CHX} = \frac{2}{\pi} A_r B_r c_p \rho_m u_{1,CHX} \tag{3.15}$$

where $B_r$ is the porosity of stack (and heat exchangers) and $u_1$ is the amplitude of oscillatory velocity at the heat exchanger location.

The pressure oscillations in the resonator are adiabatic and hence, the principal mode of heat transfer in the resonator (neglecting the thermoacoustic effect at the resonator walls) is heat diffusion.

A typical control volume of the stack region is shown below in Figure 3.4. The temperature variation of this control volume can be determined by the heat flowing in and out of it.

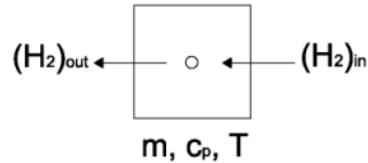

**Figure 3.4** Heat flowing in and out of a cell of 'stack' region.

The variation is given by:

$$mc_p \frac{dT}{dt} = (H_2)_{in} - (H_2)_{out} \quad (3.16)$$

This one-dimensional unsteady state heat transport equation governs the temperature distribution in the stack in transient regime.

### 3.4.4 Computational Domain and Boundary Conditions

The one-dimensional computational domain consists of the hot heat exchanger, the stack, the cold heat exchanger and the gas column in the resonator in order as shown above in Figure 3.3. Thus, the domain is bounded by hot heat exchanger at the left end and by rigid termination at the extreme right end of the resonator.

An isothermal boundary condition is imposed at the hot heat exchanger. In practical case, such a boundary condition can be realized by maintaining its temperature at a chosen constant value by removing heat from it (for example, by circulating cooling water). The other extreme end of the domain is taken to be insulated.

### 3.4.5 Solution Methodology

The transient state equation is solved using Implicit Finite Different Method. To solve the problem, the computational domain is divided into a one-dimensional grid. Stack and the gas column inside the resonator are each divided into 'n' grid-points. In standing wave TARs, the optimal length of heat exchangers is twice the oscillatory displacement amplitude of gas at the heat exchanger location [1]. As this amplitude is quite small, the heat exchangers are modeled as 'lumped' elements. The grid representation of the computational domain is shown below in Figure 3.5.

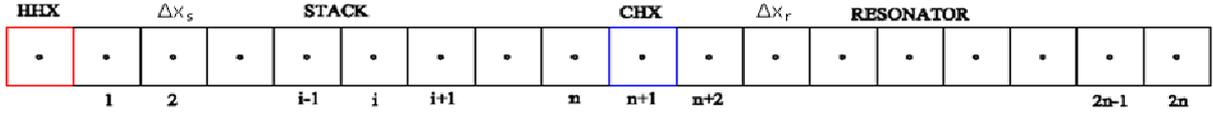

**Figure 3.5** Grid representation of computational domain.

For a general stack cell 'i', the heat balance is given by:

$$C_i \frac{dT_i}{dt} = A_{i+1}\left(\frac{dT}{dx}\right)_{i+1} + B_{i+1} - A_i\left(\frac{dT}{dx}\right)_i - B_i \tag{3.17}$$

where 'C' is the 'mc$_p$' of cell, given by:

$$C = \Pi(\rho_m c_p y_0 + \rho_s c_{ps} l)\Delta x_s \tag{3.18}$$

Here, the coefficients A and B, as well as the temperature gradients are evaluated at the cell faces. Using central difference scheme to approximate the temperature gradient and rearranging terms, eqn(3.17) simplifies to:

$$\left(-\frac{A_i}{\Delta x_s}\right)T_{i-1}^{k+1} + \left(\frac{C_i}{\Delta t} + \frac{A_i + A_{i+1}}{\Delta x_s}\right)T_i^{k+1} + \left(-\frac{A_{i+1}}{\Delta x_s}\right)T_{i+1}^{k+1} = \left(\frac{C_i}{\Delta t}T_i^k + B_{i+1} - B_i\right) \tag{3.19}$$

The finite difference equation for the cold heat exchanger ('n+1'th cell) is:

$$\left(-\frac{kA_r}{\Delta x_{res}}\right)T_n^{k+1} + \left(\frac{C_{n+1}}{\Delta t} + h_{CHX} + \frac{kA_r}{\Delta x_r}\right)T_{n+1}^{k+1} + \left(-\frac{kA_r}{\Delta x_{res}}\right)T_{n+2}^{k+1} = \left(\frac{C_{n+1}}{\Delta t}T_{n+1}^k\right) \tag{3.20}$$

Similarly, for an 'i'$^{th}$ cell of hollow resonator, the discretised equation is given by:

$$\left(-\frac{kA_r}{\Delta x_{res}}\right)T_{i-1}^{k+1} + \left(\frac{C_i}{\Delta t} + \frac{2kA_r}{\Delta x_{res}}\right)T_i^{k+1} + \left(-\frac{kA_r}{\Delta x_{res}}\right)T_{i+1}^{k+1} = \left(\frac{C_i}{\Delta t}T_i^k\right) \tag{3.21}$$

The resultant system gives temperature of all the cells constituting the computational domain at time step 'k+1' in terms of temperatures at previous time step 'k'.

Eqns(3.19 – 3.21) represent a system of '2n' linear equations in '2n' unknowns. The system can be written in a matrix form:

$$a \times T = b \qquad (3.22)$$

where 'a' is a 2n x 2n square matrix comprising of the coefficients of temperatures of 'k+1'$^{th}$ time step (*i.e* LHS), 'T' is 2n x 1 column matrix representing temperatures of 'k+1'$^{th}$ time step and 'b' is again 2n x 1 column matrix with the terms on RHS of the system of equations.

The Gauss inversion scheme is used to calculate temperatures of 'k+1'th time step. The temperatures of old time step 'k', are then updated by the calculated values of new time step. This procedure is iteratively repeated till the temperatures at two successive time steps do not differ significantly.

The transient state temperature profiles, the cooldown curves and other theoretical results of this analysis are presented in a later chapter.

## 3.5 Operating Parameters and Working Gas

This section describes the implications of the operating parameters, the working gas and the stack material on the TAR performance.

### 3.5.1 Operating Parameters

The important operating parameters that govern the performance of a TAR are the mean pressure and the mean temperature of the working gas, the operating frequency and the dynamic pressure. The energy density of an acoustic field is directly proportional to the mean pressure of the medium and the operating frequency. Hence, it is advisable to keep both of these as high as possible. However, the maximum allowable mean pressure is limited by the strength of the TAR hardware material. It can also be noticed from the expression of thermal penetration depth that, it becomes lesser with high mean pressure and operating frequency. As a result, less amount of gas remains in thermal contact with the plates of a given stack, thereby decreasing the heat pumping effect. Hence, while deciding the mean pressure and operating frequency, a compromise has to be made between be made between device size and its heat pumping capacity.

The other parameter on which the performance of a TAR depends, is the dynamic pressure. The cooling power, which a TAR can generate at the cold end, is directly proportional to square of the dynamic pressure amplitude. Hence, a large dynamic pressure is preferable. However, beyond a certain upper limit of dynamic pressure, turbulence effects set in and hamper the TAR performance. The transition to the turbulent or the non-linear regime of operation is governed by the boundary layer Reynold's Number, given by [1]:

$$\text{Re}_a = \frac{\rho_m u_1 \delta_v}{\mu_m} \qquad (3.23)$$

In order to prevent the non-linear effects, the boundary layer Reynold's Number should be less than 500 [Swift paper]. In practice, the dynamic pressure in a standing wave TAR is limited to about 2-3 % of the mean pressure [3,16].

In a loudspeaker driven standing wave TAR, the obtainable dynamic pressure in the resonator depends on the mean pressure and the operating frequency. The dynamic pressure amplitude has a maximum value when the TAR is operated at resonance. Similarly, at higher mean pressure, the dynamic pressure is also higher. The effect of mean pressure and resonance frequency on dynamic pressure is described in detail in a later chapter.

### 3.5.2 Working Gas

The choice of working gas is as important as the choice of operating parameters. The working gas should have a high sound speed so that the energy density of acoustic field for a given operating frequency is high. This helps to reduce the system size for a certain cooling requirement. However, the overall system length increases with the sound speed. Other important thermodynamic properties of the working gas are its thermal conductivity, specific heat capacity ratio and viscosity. It is advisable to use a gas with lower thermal conductivity so that the loss of cooling power due to axial conduction in the stack is low. But, lower thermal conductivity results in lower thermal penetration depth which in turn is detrimental to the cooling power. Higher specific heat capacity ratio results in higher temperature oscillations (eqn(1.1)) in the gas for same magnitude of dynamic pressure. This helps to improve the heat pumping capacity of the TAR. The working gas should have a low viscosity so as to minimize the loss of acoustic power in the viscous penetration depth. In addition to all these, the working gas should be inert from the point of view of safety of life, environment friendly, readily available and cheap.

### 3.5.3 Stack Material

The choice of material for constructing the stack is based in its properties like thermal conductivity, specific heat capacity and density. Low thermal conductivity ensures less cooling power loss due to axial conduction through the stack. Its specific heat capacity should be higher than that of working gas so that its temperature fall remains steady.

### 3.5.4 Design Choices

Helium was chosen as the working gas for designing the TAR. This is because Helium has the highest thermal conductivity and sound speed among inert gases. Mylar was chosen as the material for making the stack due to its low thermal conductivity and ready availability at a cheap price. The choice of mean pressure is done by theoretical parametric analysis, while the operating frequency is chosen so as keep the resonator length within reasonable limits. The operating parameters, properties of working gas and the stack material are given below in Table 3.1.

**Table 3.1** Operating Parameters, Working Gas and Stack Material Properties

| Operating Parameters | | Working Gas Properties (@ $T_m$) | | Stack Material Properties (@ $T_m$) | |
|---|---|---|---|---|---|
| $p_m$ | 10 bar | k | 0.152 W/m-K | $k_s$ | 0.16 W/m-K |
| f | 400 Hz | $c_p$ | 5193.4 J/kg-K | $c_s$ | 1110 J/kg-K |
| $T_m$ | 308 K | a | 1020 m/s | $\rho_s$ | 1347.5 kg/m$^3$ |
| | | μ | 1.99E-5 kg/m-s | | |
| | | σ | 0.679 | | |

At the above mentioned operating parameters, the values of thermal penetration depth and the viscous penetration depths in Helium are 1.2E-4 m and 9.9E-5 m respectively.

## 3.6 Geometric Dimensions

The dimensions of the TAR derived from the theoretical analysis are given in this section.

### 3.6.1 Resonators for Dynamic Pressure Measurements

The lengths of straight resonators used for these measurements are derived from the acoustic impedance analysis (section 3.2.1). To investigate the effect of working gas on dynamic pressure, two resonators were designed. From eqn(3.5), the lengths of Helium resonator and Nitrogen resonator respectively are calculated as 1250 mm and 450 mm. These lengths ensured a half wavelength standing wave in each case at 400 Hz. Because of ready availability in market, the inner diameter of the resonators was chosen to be 32 mm. It is to be noted that for a straight resonator of uniform cross section, the resonance frequency is independent of the diameter.

### 3.6.2 Resonator of TAR

The resonator for TAR has a shape as shown above in Figure 3.2 and its dimensions were determined using DeltaEC. The dimensions are given below in Table 3.2 below:

**Table 3.2** Dimensions of TAR resonator

| Length | Value | Diameter | Value |
|---|---|---|---|
| L1 | 0.1 m | D1 | 0.032 m |
| L2 | 0.02 m | D3 | 0.0126 m |
| L3 | 0.1846 m | D4 | 0.12 m |
| L3 | 0.1 m | | |

The total length of the resonator was 0.446 m and which corresponds to a quarter wavelength standing wave in Helium at 400 Hz.

### 3.6.3 Stack

For ease of construction, spiral geometry was chosen for the stack. The Mylar film thickness of 0.18 mm was chosen because of its ready availability. The spacing between layers of the stack was taken to be 0.3 mm. This spacing corresponds to 2.5 δk which is close to its optimal value of 2 δk. The porosity of the stack was 0.53. The length of the stack was chosen to be 100 mm and thickness of the roll to be 32 mm so as to fit tightly into the larger diameter part of TAR resonator.

The fabrication of all the TAR components is described in detail in the following chapter.

# Chapter 4

# FABRICATION AND EXPERIMENTAL SETUP

## 4.1 Introduction

This chapter describes the fabrication of the various TAR components. Several aspects related to fabrication like the choice of materials, the machining processes involved, etc. are addressed. The fabrication and assembly of the two sub-systems of a TAR – the acoustic driver assembly and the refrigerating assembly is described in separate sections. Fabrication details of the setup for dynamic pressure measurement are also given. The chapter concludes with the description of experimental setup and the allied instrumentation.

## 4.2 Acoustic Driver

The acoustic driver constructed for the present investigations is based on an electro-dynamic motor of a moving coil loudspeaker. Its main components are – the magnet and the pole piece, the voice coil and the sound radiating cone, the suspensions and mechanical vibration dampers, and the supporting rings. This assembly is mounted on a flange and is enclosed by a cylindrical back jacket. The schematic of the acoustic driver with its various components is shown below in Figure 4.1.

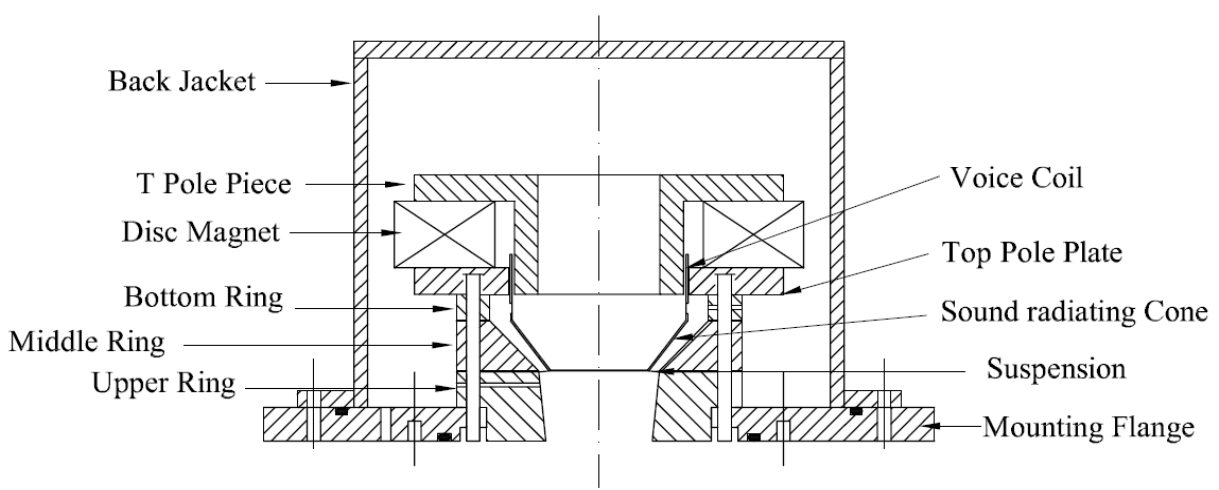

**Figure 4.1** Schematic of the acoustic driver assembly showing its various components

## 4.2.1 Fabrication

### a) Magnet - Pole Piece

For the construction of the acoustic driver, the magnet and the pole pieces of a commercially available moving coil loudspeaker are used. The magnet-pole piece assembly was carefully taken out as such from the loudspeaker. The assembly consisted of a ferrite disc magnet sandwiched between a top pole plate and a 'T' pole piece. Both the pole pieces were made of soft iron. The schematic of the magnet pole-piece assembly is shown in Figure 4.2

The ferrite disc magnet is 20 mm thick with inner diameter (ID) 60 mm and outer diameter (OD) 120 mm. The top pole plate had an ID 53.6 mm and an OD 110 mm. Its thickness is 8 mm which is the height of the flux gap. The diameter of the stem of the 'T' pole piece is 50.5 mm. Hence, the radial width of the flux gap is 1.56 mm. The total height of the 'T' piece is 36 mm. The 'T' piece has a bore of diameter 36 mm in the stem so as to vent out the heat produced during operation of the speaker.

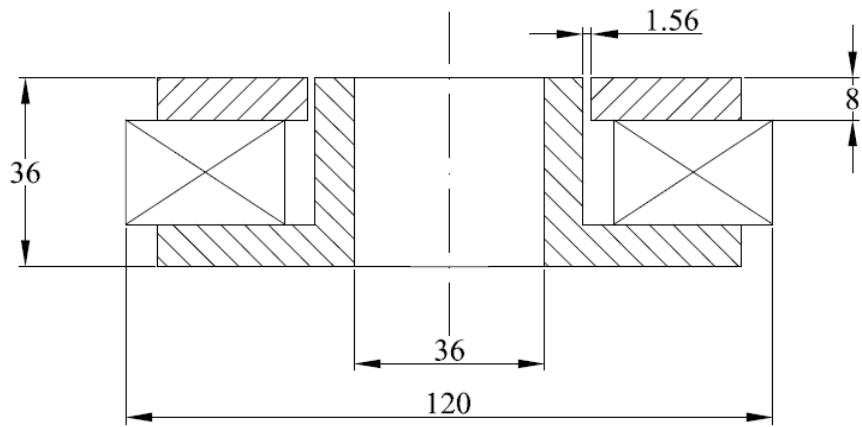

**Figure 4.2** Ferrite disc magnet and soft iron pole pieces

The magnet – pole piece generated a radial magnetic field of strength 0.99 T in the flux gap.

### b) Voice Coil and Sound Radiating Cone

In the present work, two different voice coils are used. The first voice coil came along with the loudspeaker. This voice coil consisted of a 34 gauge super insulated copper wire (diameter 0.26 mm) wound on a thin Kapton film covered with a black craft paper. The voice coil is two layered and its diameter is 52 mm. The total winding height is 14 mm with a static overhung of 3 mm on both the sides of active coil height. Thus, the active coil height is 8 mm (equal to the height of the top pole plate). A thin walled Aluminum cone is glued to this voice coil. The conical shape provides the necessary reduction of cross section from 52 mm voice coil diameter to 32 mm resonator diameter. This thickness of the Aluminum cone is 1 mm in order to ensure low weight. This thickness was limited due to constraints in the in-house manufacturing operation. The height of the cone is 17 mm. This coil when suspended in the magnetic flux gap, yields a force factor (Bl) close to 9.8 T-m. The DC resistance of the coil is 6.8 Ω while its inductance is 0.103 mH. The total mass of this voice-coil and reducer cone is about 23 g.

In order to reduce the excess moving mass and DC resistance of the voice coil due to large overhung, a new voice coil is designed. A thin walled Delrin coil former is manufactured for this purpose. Copper wire of same gauge as above is used for winding. The coil is wound in-house. The new voice coil has the same diameter as the old one but the winding height is only 10 mm. Thus, the overhung is reduced to just 1 mm on each end of the flux gap height. This brought down the total mass to just 12 g. The DC resistance is also reduced to 4 Ω.

The schematics of the two voice coils are shown below in Figure 4.3a and Figure 4.3b

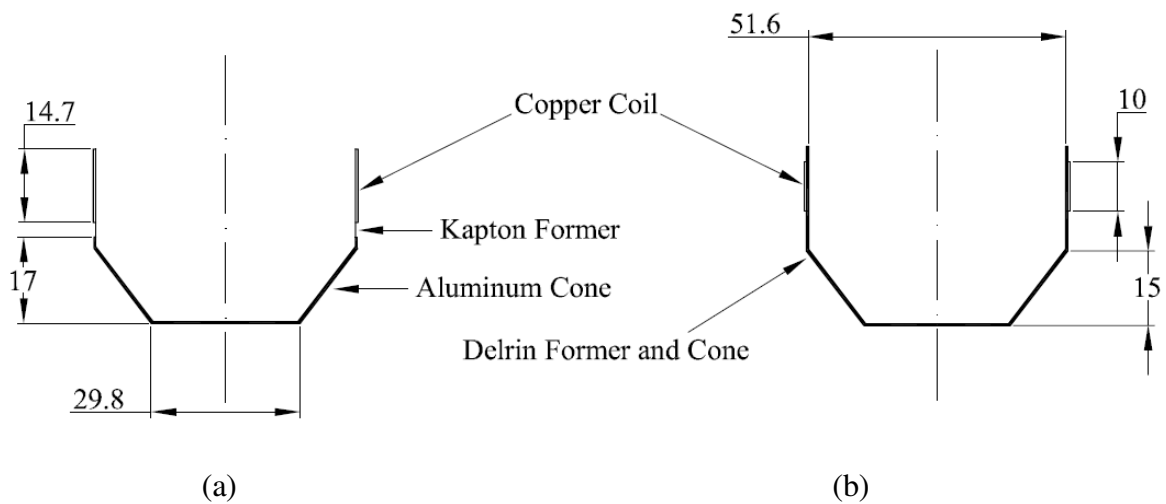

(a)            (b)

**Figure 4.3** The voice coils used for present work (a) Old voice coil (b) New voice coil

c) **Supporting Rings**

Three rings and a suspension are used to suspend the voice coil in the magnetic field. The bottom ring is placed on the top plate of the magnet system and it forms the base of the coil support structure. Its placement on the top pole plate is very crucial so as to ensure that the suspended voice coil sits uniformly in the narrow magnetic flux gap. This is ensured by placing the bottom ring exactly concentric to the circular flux gap. The ID and the OD of the bottom ring are respectively 67 mm and 85 mm, while its height is 10 mm. A 2 mm diameter hole drilled radially through the bottom ring provides passage for the electrical connections to the voice coil. The bottom ring is made of Aluminum.

The middle ring has a cylindrical shape from the outside. The inside of it is conical, which encloses the conical section of the voice coil. The OD of this ring is 85 mm. The inner conical surface provides reduction of diameter from 67 mm to 36 mm. The total height is 15 mm. This component is also made of Aluminum.

The top ring facilitates the attachment of the magnet voice-coil system onto the mounting flange. It has a 'T' shape with the stem fitting in the mounting flange. The inner bore in this ring is slightly conical, reducing the diameter from 36 mm to 32 mm over a length of 20.7 mm. A 0.5 mm diameter capillary is passed through the stem radially. This capillary ensures equal static pressure in front and back of the voice coil. The top ring is made of Aluminum.

All of the supporting rings carry six M4 holes on a 76 PCD. The schematics of the support rings are shown below in Figure 4.4a, Figure 4.4b and Figure 4.4c.

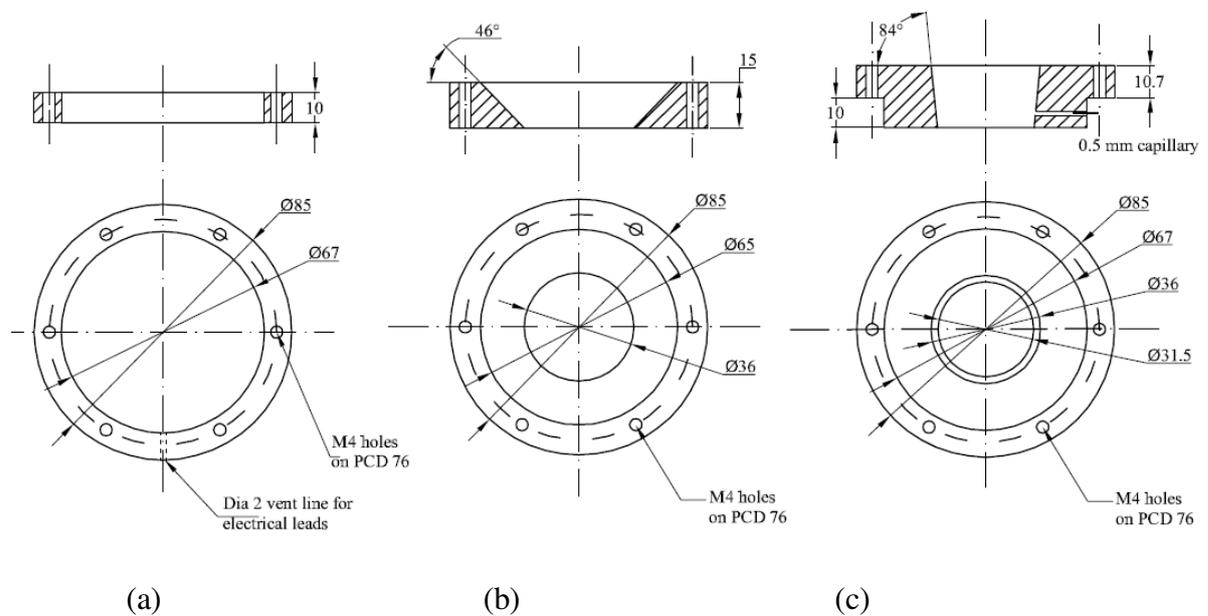

(a)  (b)  (c)

**Figure 4.4** Schematics of the (a) Bottom, (b) Middle and (c) Top supporting rings

**d) Mounting Flange**

The mounting flange is the interface between the driver assembly and the refrigerating assembly. The driver assembly and the resonator are mounted on two opposite faces of the mounting flange. For its strength requirements to be able to hold pressurized gas in the back volume, the mounting flange is made out of SS 304. The overall diameter of the mounting flange is 220 mm and its thickness is 10 mm. The bore at its center is stepped. It has a diameter of 65 mm on driver side and 83 mm on resonator side. Six M4 holes on 76 PCD facilitate attachment of the driver assembly. Six M4 tapped blind holes on 110 PCD provide for mounting the resonator. Appropriate grooves for placing gas sealing O-rings are made on the mounting flange. The schematic of the mounting flange is shown below in Figure 4.5.

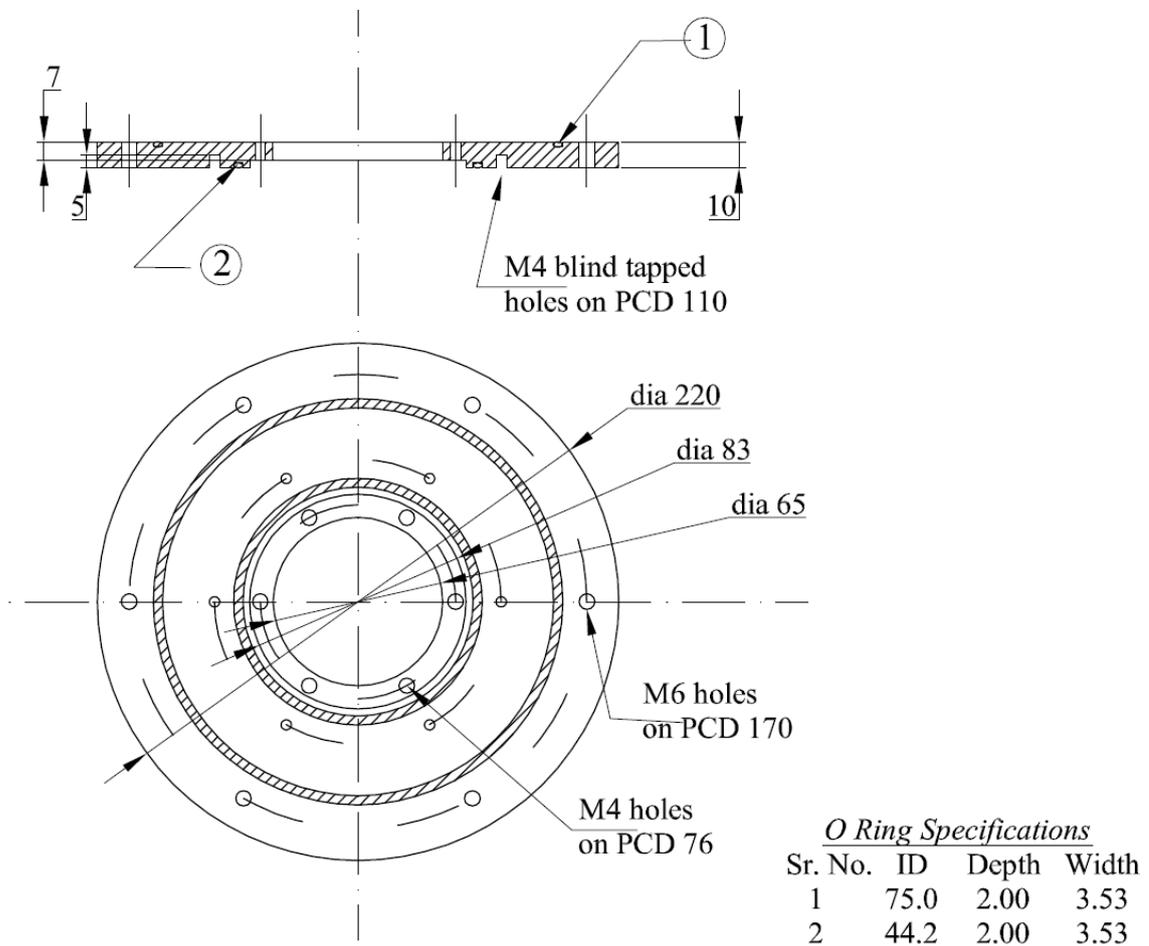

**Figure 4.5** Schematic of the Mounting Flange

**e) Back Jacket**

The back jacket encloses the acoustic driver assembly. It consists of a cylindrical jacket with a flat plate welded on its one end. A flange is welded along its periphery of its other end so as to attach the back jacket to the mounting flange. To ensure high strength, the back jacket is made of SS 304 and all the joints are prepared by means of argon welding.

The overall height of the jacket is 55 mm and the ID of its cylindrical body is 138 mm. The body is 4 mm thick. The back jacket also carries a gas charging line, a provision for attaching a static pressure gauge and a 2 pin electrical feed-through for powering the acoustic driver. The schematic of the back jacket is shown below in Figure 4.6.

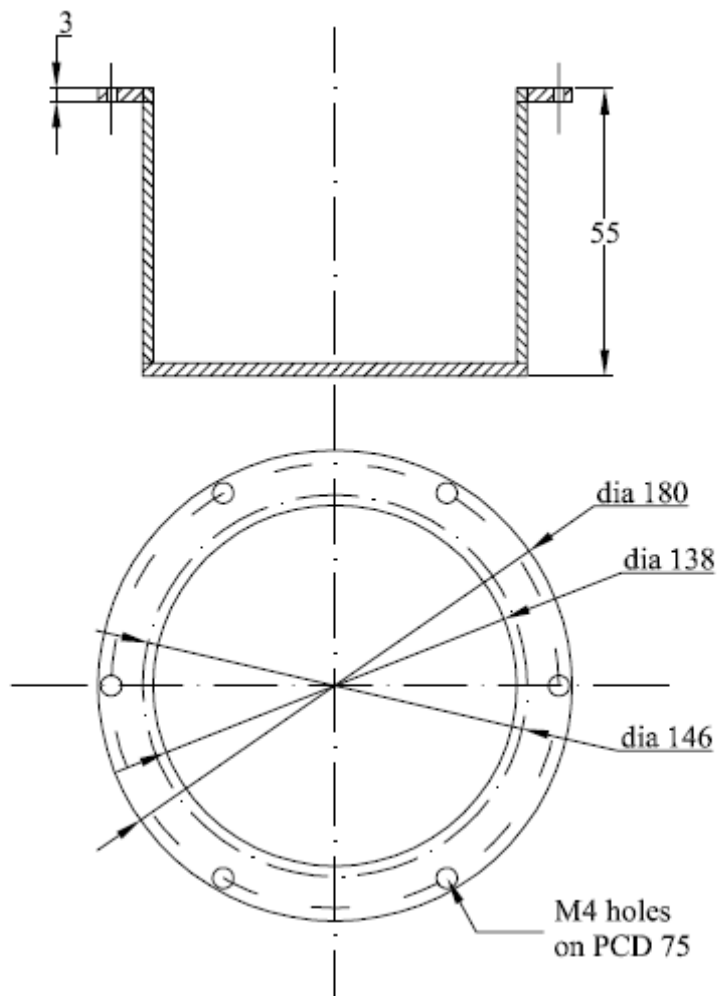

**Figure 4.6** Schematic of the Back Jacket

## 4.2.2 Assembly

After ensuring proper connectivity of the voice coil, the first step in the assembly is attaching the suspension to the coil. In the present case, a 0.7 mm thick neoprene rubber sheet is used as the suspension. The circle is cut from the rubber sheet and is glued onto the middle ring. The conical termination of the voice coil is then glued to this circle. This procedure can be understood better from Figure 4.7a and Figure 4.7b as shown below:

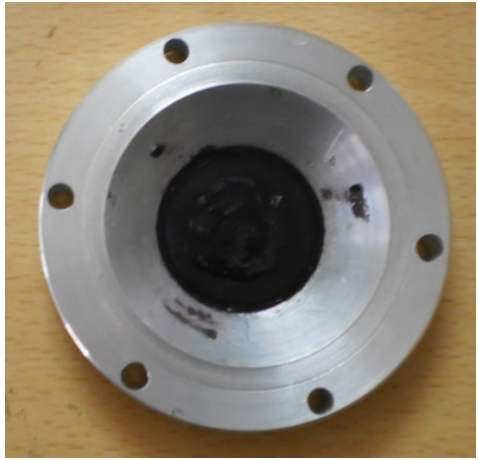 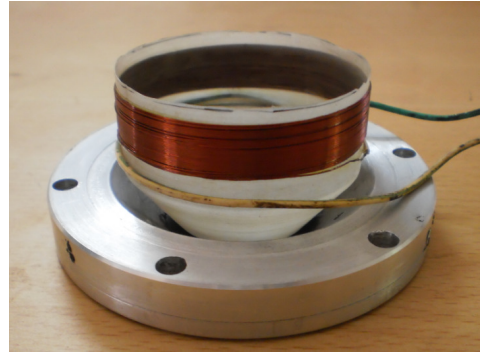

(a)                                                                 (b)

**Figure 4.7** Attaching voice coil to the suspension
(a) Rubber Suspension is glued to the middle ring (b) Then the Voice Coil is glued to the suspension

To mount this voice coil in the magnetic field, three long bolts are attached to the top plate of magnet assembly as shown in Figure 4.8a. The bottom ring is first mounted on the bolts. This is followed by the middle ring and the voice coil, and the top ring. The electrical wires from the voice coil are taken out through a hole in the bottom ring. This assembly is shown in Figure 4.8b. In order to damp the mechanical vibrations during operation, a rubber sheet is placed between each of the three rings and the top plate of magnet assembly.

This assembly is then bolted onto the mounting flange and is enclosed by the back jacket. The driver assembly is thus ready for operation.

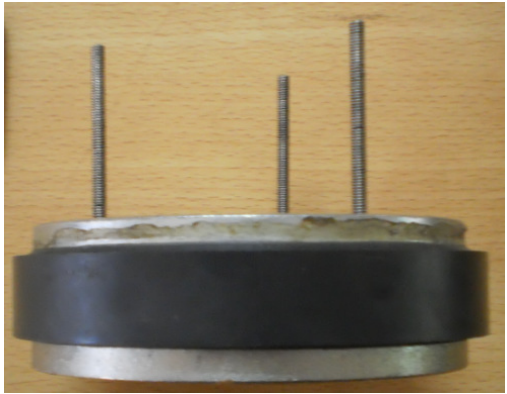
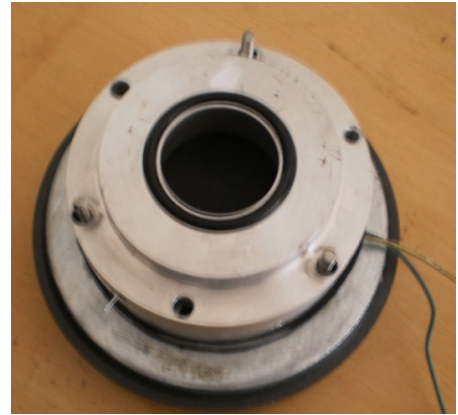

(a)                              (b)

**Figure 4.8** Suspending the voice coil in magnetic flux gap
(a) Long bolts attached to the top plate of magnet system. (b) Ready assembly

## 4.3 Resonators for Dynamic Pressure Measurement

Straight hollow pipes were used as resonators for this investigation. SS 304 pipes of 32 mm ID and 38 mm OD are available as standard. A 4 mm thick SS 304 flange is argon welded to seal of one of the ends of the resonator pipe. Another flange with a provision for mounting on the driver is welded at the other end on the periphery. Two resonators of lengths 1250 mm and 450 mm were constructed. These lengths correspond to the half wavelengths at 400 Hz in Helium and Nitrogen gases respectively. During operation, the resonator was bolted onto the mounting flange and the junction was sealed by means of a rubber O-ring.

## 4.4 Refrigerator Assembly

The refrigeration assembly can be divided into three sub-groups- the warm HX, the vacuum chamber and the mounting flange, and the resonator system. The warm heat exchanger and mounting flange group consists of a warm heat exchanger which facilitates heat exchanger between the working gas and the ambient. It also has a mounting flange and a provision for water cooling. The resonator system comprises of the stack, the stack holder, the cold heat exchanger (CHX), resonator tube and the buffer volume. The resonator system is designed in DeltaEC so as to have quarter wave resonance in Helium at 400 Hz. The schematic of the refrigerating assembly is shown in Figure 4.9.

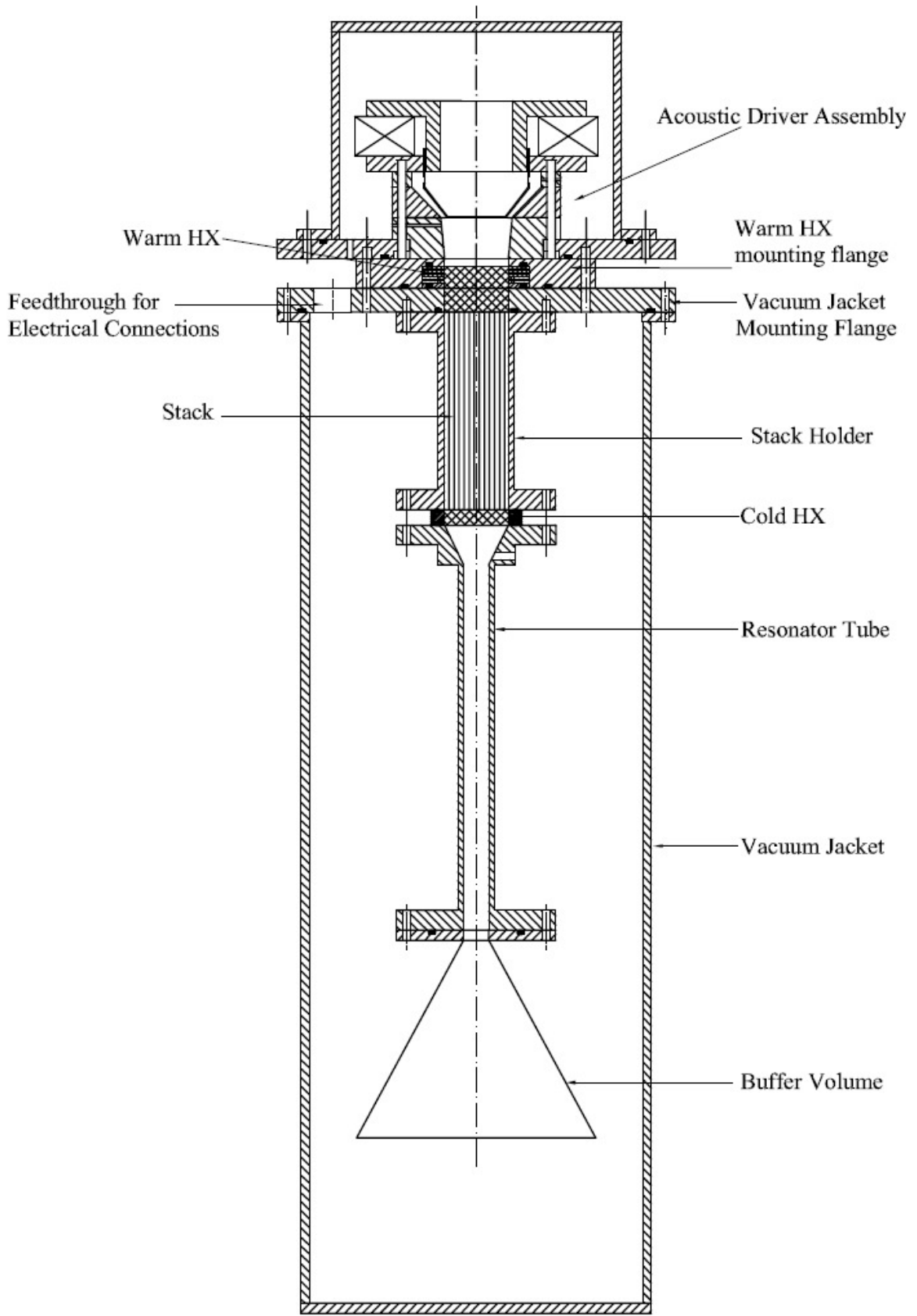

**Figure 4.9** Schematic of the refrigeration assembly showing various sub-groups are classified above

## 4.4.1 Fabrication

**a) The Spiral Stack**

The stack is manufactured from a 0.18 mm thick Mylar film. 0.3 mm thick Nylon fishing lines are used as spacers. As discussed in the previous chapter, spiral geometry is chosen for the stack, because of its ease in manufacturing. The distance between two adjacent spacing lines is 5 mm throughout the stack cross section. This particular spacing ensured that the two layers of the Mylar film do not touch each other and the gas passage channels are uniform. This has been realized through repeated attempts of making the stack. The length of the stack is 100 mm and its diameter is 32 mm. The stack manufacturing process is described as follows: A long wooden plank wide enough to accommodate the width of Mylar film is taken and equidistant slits at 5 mm from each other are made on both its edges. The Mylar film is then held tightly on the plank. Nylon fishing line is wound over the Mylar film. This can be visualized better from Figure 4.10. After each turn, the fishing line passes through the next slit which is as 5 mm from the previous one. This ensures a winding pitch of 5 mm and hence, a spacing of 5 mm between two consecutive fishing lines. The fishing lines are then glued to the Mylar film. This can be either done by means of an adhesive tape or an adhesive like insulating varnish. It is found that the use of adhesive tape increases the thickness of the Mylar film layer and hence, use of insulating varnish is advisable. The fishing lines are then cut along the edges of the Mylar film. Thus, the stack is ready for rolling. Rolling is carefully done so as not to disturb the spacing lines.

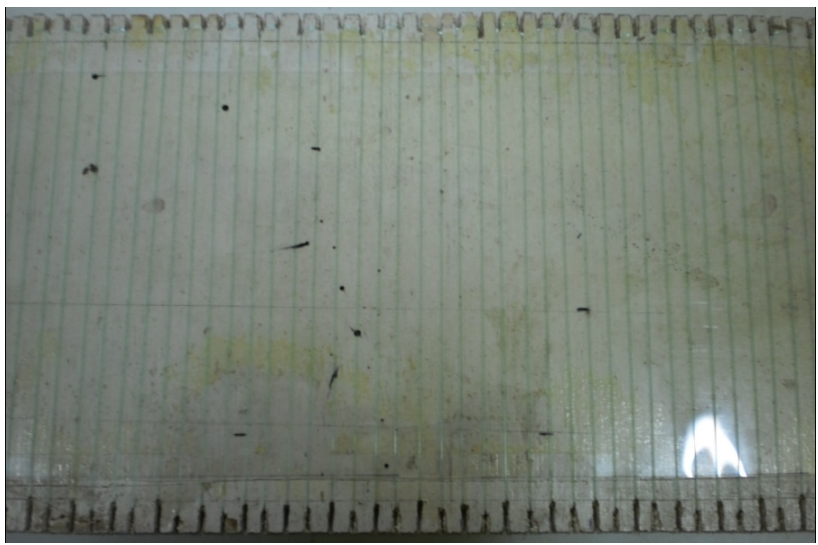

**Figure 4.10** The wooden plank with slits on its edges. The Mylar film and Nylon fishing lines can also be seen.

**b) The Stack Holder**

The stack holder houses the stack and is sandwiched between the two heat exchangers. The stack material should have a low thermal conductivity to minimize the axial conduction losses when a temperature difference is generated across the stack. It should be strong enough to hold the gas at the desired charging pressure. To meet these requirements, Delrin was chosen as the material for the stack. Apart from being light weight ($\rho$ = 1420 kg/m$^3$) and poor conductor of heat (k = 0.3 W/m-k), it has the best machinability among other polymers like PTFE, nylon, etc.

The total length of the stack holder is 100 mm and its inner diameter is 32 mm. It has a bolting flange of diameter 80 mm and thickness 10 mm at each of its end. The entire structure is machined out as an integral part from a Delrin rod. The schematic of the stack holder is shown in Figure 4.11,

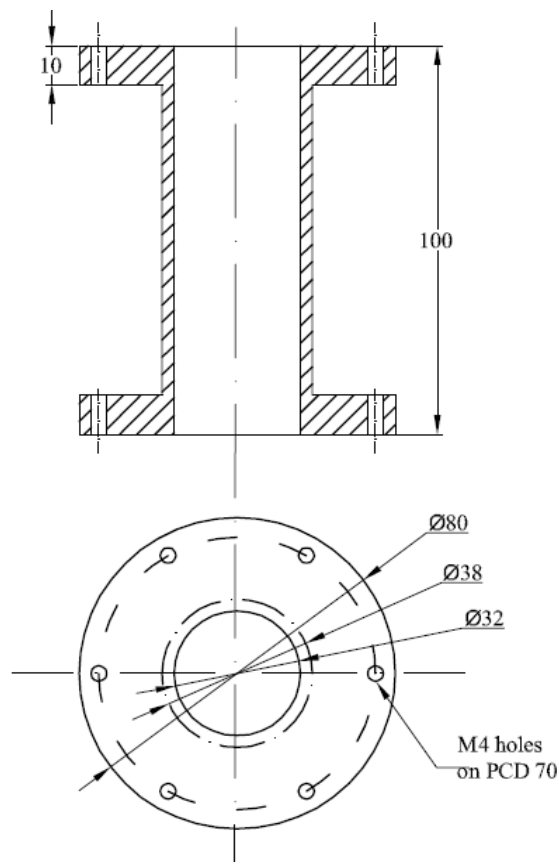

**Figure 4.11** Schematic of the Stack Holder

**c) Heat Exchangers (HXs)**

The HXs facilitate the exchange of heat from gas near the hot/cold end of the stack and the ambient. It also provides a surface to mount the sensors for measurement of temperature. Because of the requirement of high thermal conductivity, the HXs are made out of soft copper.

The CHX has the shape of a ring of 32 mm ID and 46 mm OD. The length of the CHX was 8 mm. This length provided easy access to the CHX circular surface and enough space so as to mount a sensor for temperature measurement. However, this length is much more than the optimal HX length as given in literature [1]. The gas side of the CHX is filled with a stack of copper screens of size #100. The schematic of the CHX body is shown in Figure 4.12a.

The body of warm HX is to be maintained at ambient by circulating cooling water. Hence, to increase the heat exchange area, circular fins are provided. The ID of warm HX is 32 mm and its OD is 54 mm. Its total length is 11 mm with four circular fins of thickness 1 mm and length 8 mm. The schematic of the warm HX is shown below in Figure 4.12b.

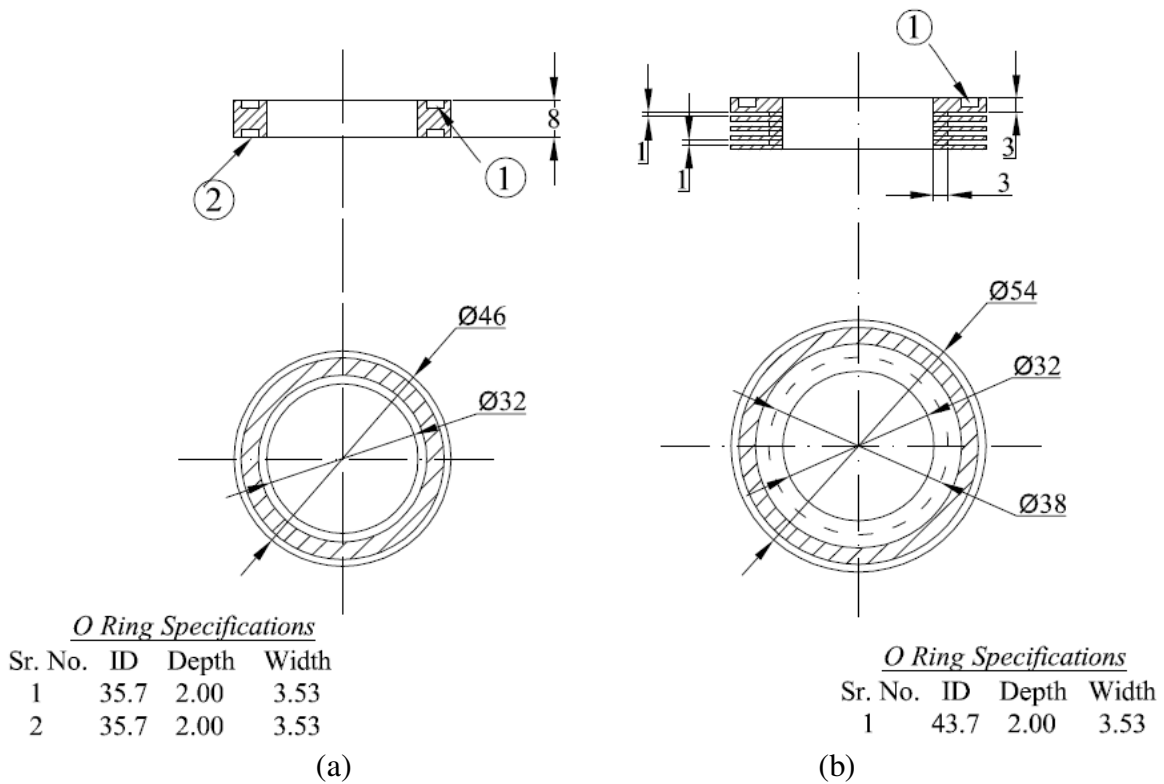

**Figure 4.12** Schematics of a) Cold HX b) Warm HX

**d) Resonator Tube**

The resonator tube is the hollow component of the resonator system. It is placed between the cold HX and the buffer volume. To ensure low thermal conductivity, high strength and light weight, the resonator tube was machined out of a Delrin rod. The resonator tube consists of a conical taper followed by a straight hollow tube. The conical taper provides the reduction of diameter from 32 mm at the cold HX side to the 12.6 mm diameter tube. The total length of the resonator tube is 205 mm. The resonator tube has flanges of diameter 80 mm on both the end for fastening to other components of the assembly. The resonator tube a tapped hole for connection of dynamic pressure transducer. The schematic of the resonator tube is shown in Figure 4.13.

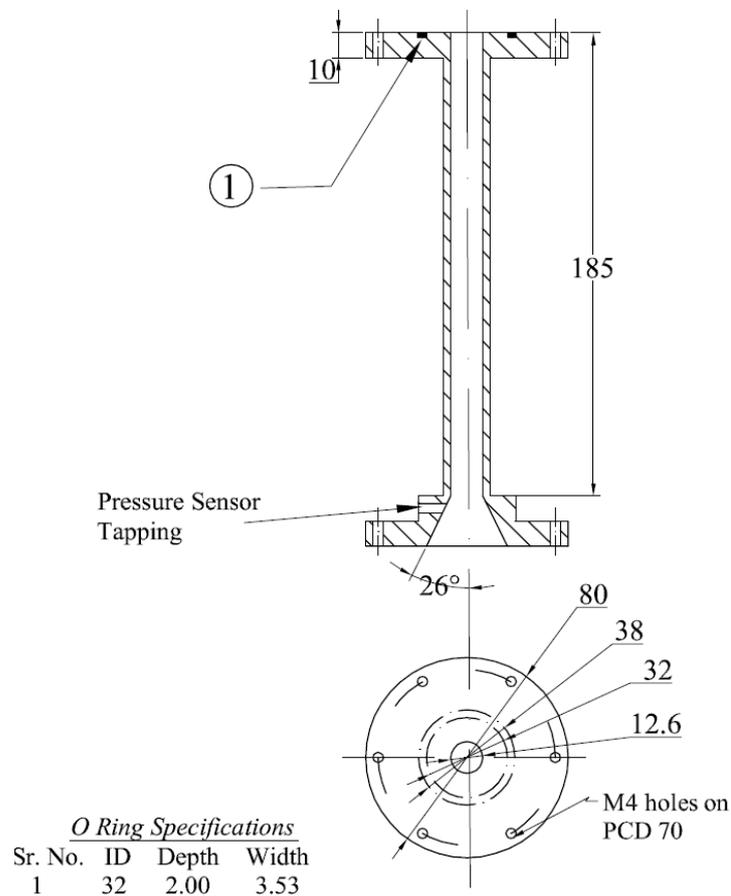

**Figure 4.13** Schematic of the Resonator Tube

**d) Buffer Volume**

The buffer volume is a large open conical volume which simulates the 'open end' of the quarter wavelength resonator. It is made of a SS 304 sheet of 1 mm thickness. The sheet is cut in the required shape and then turned into a truncated cone on a rolling machine. The cone is then welded using argon welding technique along its slant length. The larger end of the truncated cone is closed by welding a circular cut SS 304 sheet.

The total height of the buffer volume is 100 mm, the larger and smaller diameters are respectively 120 mm and 12.6 mm. A 4 mm thick flange with 80 mm diameter is welded at the smaller opening of the cone. The flange also has a 12 mm bore concentric to that of truncated the cone. The cone can be fastened to the resonator tube with the help of this connecting flange.

The schematic of the buffer volume is shown below in Figure 4.14.

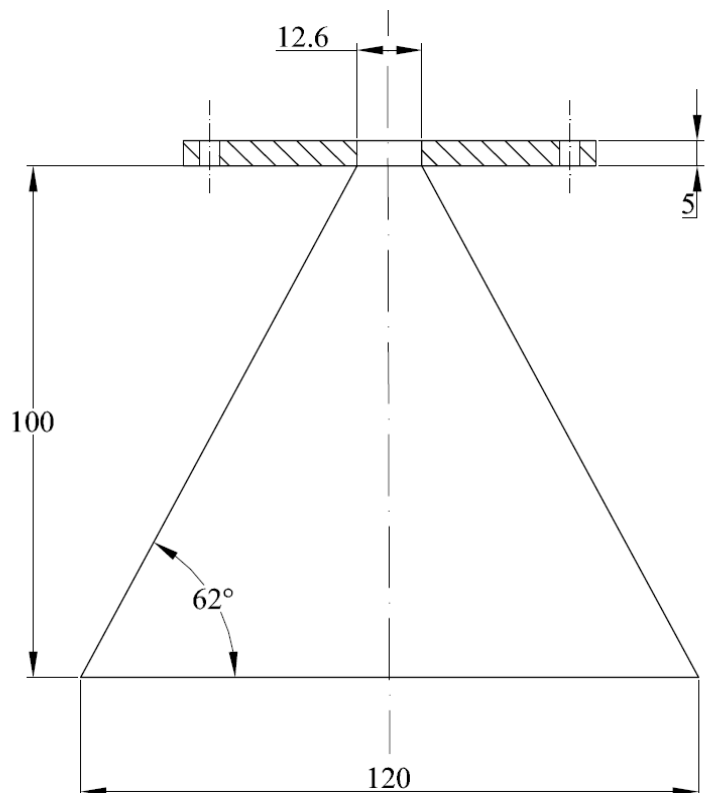

**Figure 4.14** Schematic of the conical buffer volume and the connecting flange

### e) Warm HX Mounting Flange

The warm HX mounting flange houses the warm HX block. It is made up of SS 304. It has a slot in which the warm HX blocks fits tightly. It is also equipped with channels for inflow and outflow of cooling water, to keep the warm HX at ambient temperature. The overall diameter of the warm HX mounting flange is 120 mm while its thickness is 15 mm. The schematic of this component is shown in Figure 4.15.

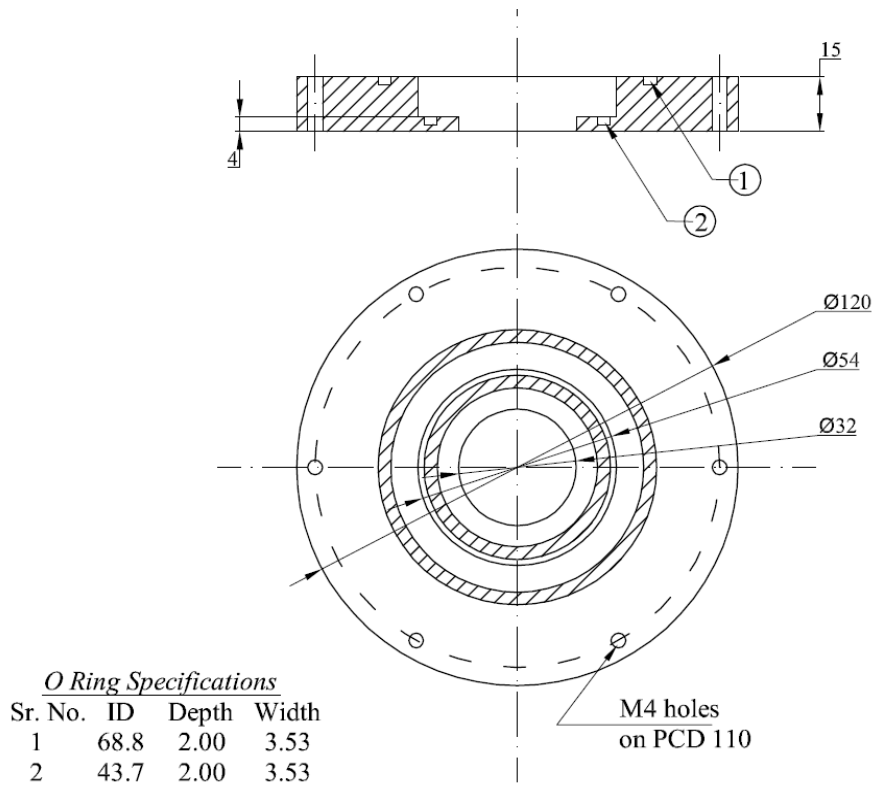

**Figure 4.15** Schematic of the warm HX mounting Flange

**e) Vacuum Jacket mounting flange**

This flange facilitates the mounting of vacuum jacket around the resonator system. It also has a provision for connecting to the driver assembly and the warm HX flange. In order to ensure light weight and high strength, the mounting flange was made of Aluminum. A nine pin electrical feed-through is connected to this flange for taking out the electrical lead wires from sensors inside the vacuum chamber.

The overall diameter of the mounting flange is 200 mm and its thickness is 12 mm. Appropriate holes and taps are made on the flange so as to fasten the resonator system and vacuum jacket on one end, and the driver assembly and warm HX flange on the other. The schematic of the vacuum jacket mounting flange is shown in Figure 4.16.

**f) Vacuum Jacket**

No vacuum jacket was manufactured due to time constraints.

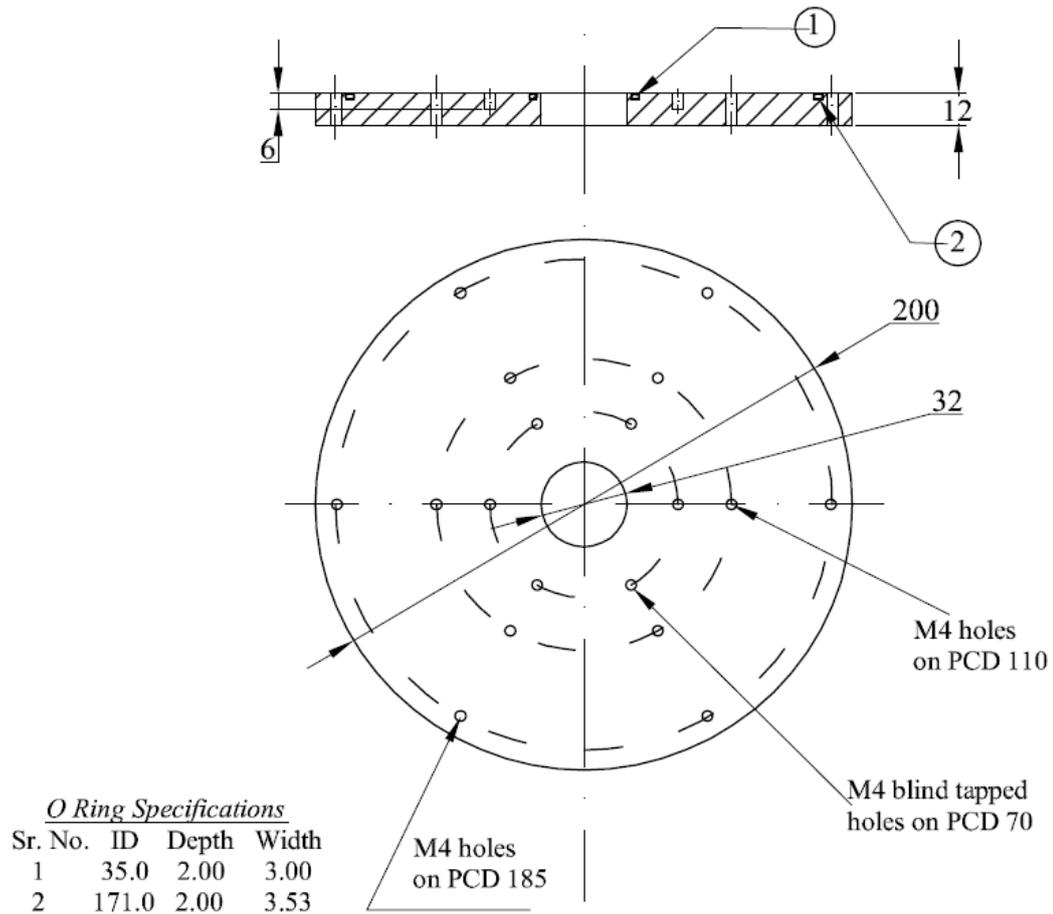

**Figure 4.16** Vacuum Jacket mounting flange

## 4.4.2 Assembly

Prior to assembly of the components, it is ensured that the stack fits tightly in the stack holder. Thereafter, the stack holder is fastened to the vacuum jacket mounting flange. The cold HX is then held between the stack holder and the resonator tube. The stack holder and the resonator tube is then fastened by means of M4 nut-bolts. Utmost care is taken while tightening these two components. As they are made of brittle Delrin, local over-tightening can simply cause the components to crack at the flange portion. To prevent this, teflon spacers are kept between the two flanges and then the nut-bolts are tightened slowly. After this, the buffer volume is fastened to the other end of the resonator tube. All the fastened joints are sealed by means of neoprene rubber O-rings.

The vacuum jacket mounting flange with the resonator system mounted, is finally fastened to the acoustic driver. The TAR is now ready for operation.

The TAR assembly along with the acoustic driver is shown in Figure 4.17

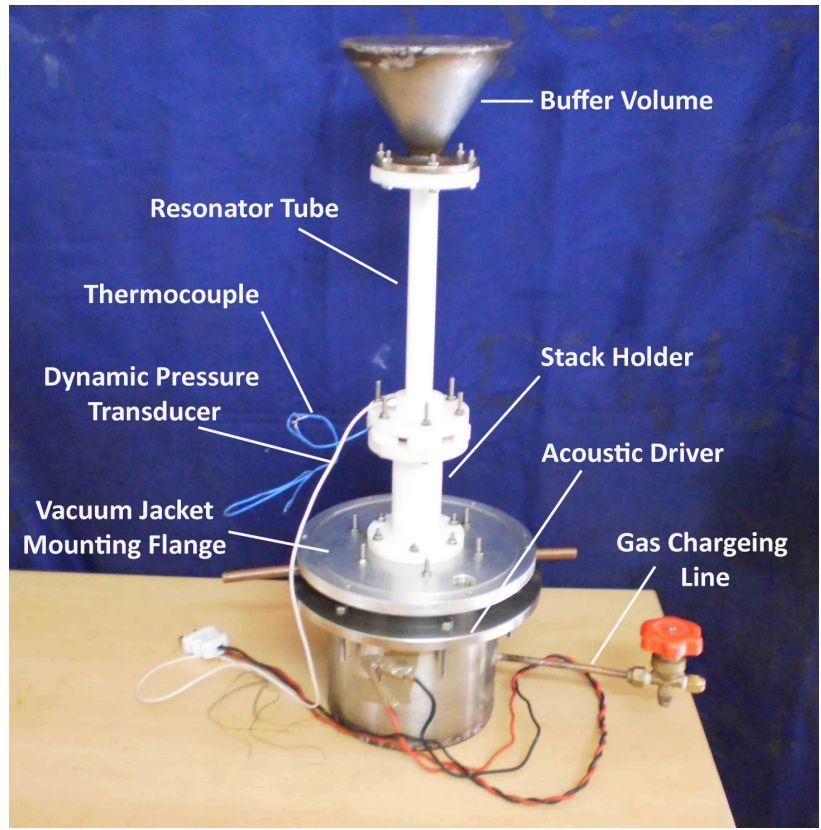

**Figure 4.17** The TAR assembly ready for operation

## 4.5 Experimental Setup and Instrumentation

The schematic of the experimental setup is shown in Figure 4.18. The TAR is powered by a variable frequency variable voltage power source. The input power, the voltage and the current are measured by means of a digital AC power meter. The magnitude of the dynamic pressure wave generated by the acoustic driver in the resonator is measured using a dynamic pressure transducer. The transducer is placed inside the resonator downstream of the stack. The small voltage signal from the pressure transducer is amplified by a differential amplifier and fed to the digital oscilloscope. The dynamic pressure is calculated from the voltage waveform observed on the digital oscilloscope. For cold temperature measurement, a copper-constantin (type-T) thermocouple is used. The thermocouple is passed through a small hole drilled through the CHX block where it is in direct contact with the cold gas inside the TAR.

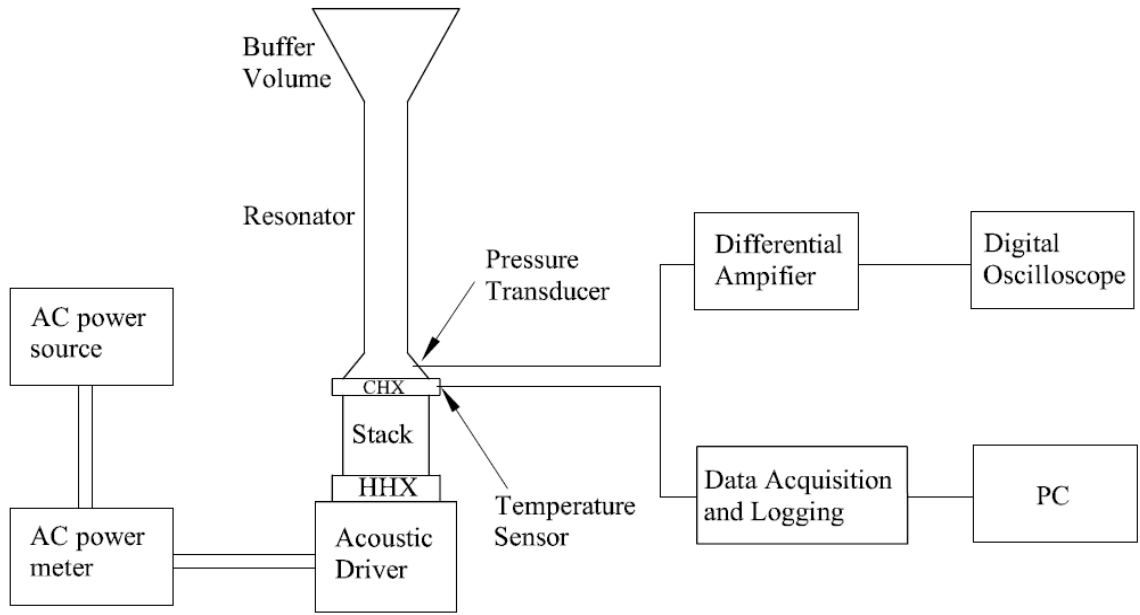

**Figure 4.18** Schematic of the Experimental Setup and Allied Instrumentation

After switching on the power supply, the temperature at the CHX of the TAR starts to fall below ambient. This fall of temperature is continuously acquired and logged by means of a DAQ system. The DAQ system is controlled by a computer. All the temperature data logged in the DAQ system is then unloaded into the PC, where it is available for post-processing. The experimental setup is shown in below in Figure 4.19

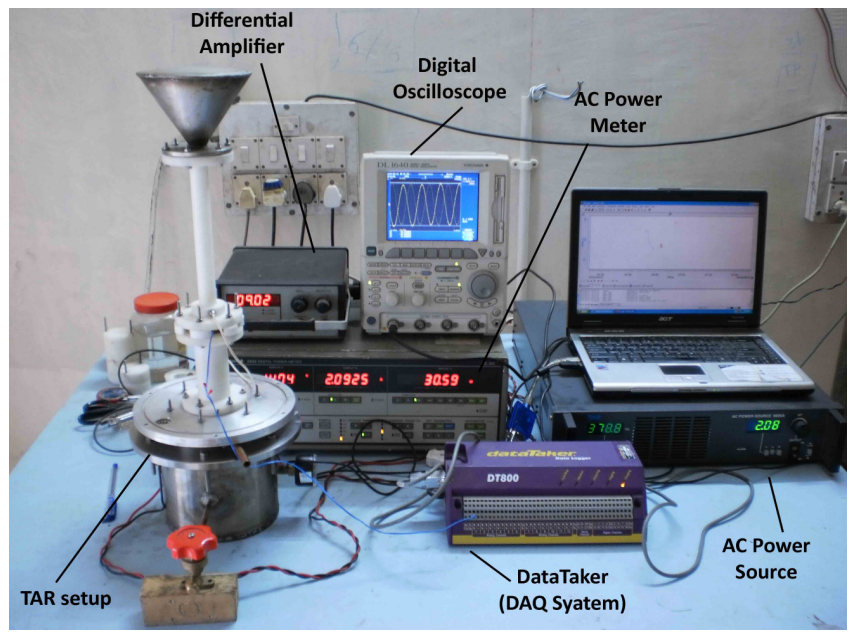

**Figure 4.19** Experimental Setup

# Chapter 5

# RESULTS AND DISCUSSION

## 5.1 Introduction

In Chapter 3, design of the quarter wave resonator for the TAR and the half wave resonators for dynamic pressure measurement has been discussed. Accordingly, these resonators and other components of the TAR setup have been fabricated. Thereafter, experimentation is carried out to determine the unknown driver parameters. The experiments for measurement of dynamic pressure in hollow pipes and the cooldown measurements of the stack are done. Transient analysis is carried out for the developed TAR setup.

This chapter presents the theoretical results of the transient state analysis. The theoretical predictions of the dynamic pressure, resonance frequency, etc. are compared with the experimental measurements. The cooldown performance of the TAR setup is also presented discussed.

## 5.2 Theoretical Results of the Transient State Model

The theoretical prediction of the temporal evolution of temperature profiles in a standing wave TAR, the cooldown time, etc are shown in this section. To validate the model, the theoretically obtained steady state cold temperatures are compared with those given in literature [1,16]. The cooldown characteristics of various working fluids are also shown using the transient state model.

### 5.2.1 Comparison with Numerical Integration

Numerical integration of thermoacoustic wave equation and enthalpy flux equation (eqn.(2.31)) for a preselected set of operating parameters and stack dimensions is a simple practice carried out to predict the steady state performance of a TAR. Here, the steady state cold temperature at different cooling loads is calculated using the FDM technique. For comparison, the operating parameters and the stack dimensions are taken same as those of Hofler's TAR [16]. Figure 5.1 shows a comparison between results obtained using present model and those reported by Hofler.

The calculations are done at two different drive ratios (Dr = $p_{max}/p_{mean}$)- 1.5 % and 3 % of the mean pressure. The drive ratios are kept small enough to ensure that there is no turbulence in the refrigerator. At a drive ratio of 3 %, the value of boundary layer Reynold's Number, $Re_a$ at the center of stack is approximately equal to 50. As predicted by the Linear Theory, the steady state cooling power obtained at the cold heat exchanger is proportional to the square of drive ratio. This indicates that, at higher drive ratios the cooling capacity of a TAR is more for a given cold end temperature. Hence, the cooling load line for Dr = 3 % is much flat as compared to the cooling load line for Dr= 1.5%.

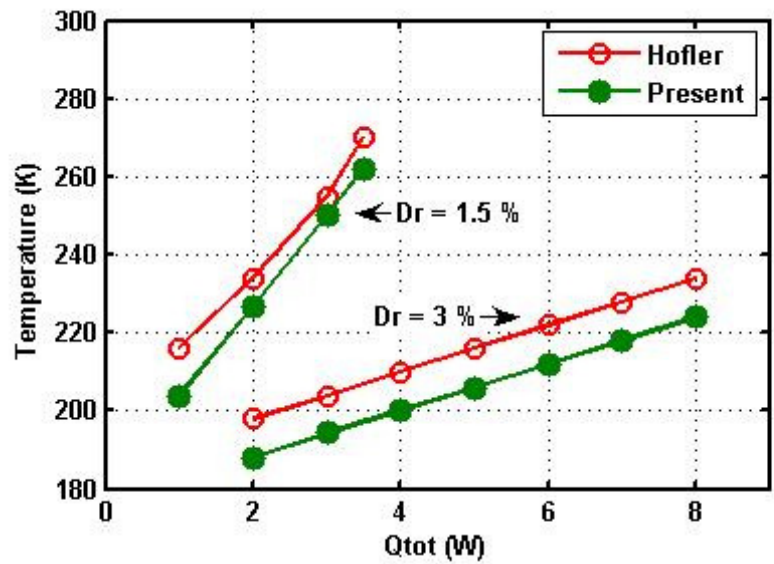

**Figure 5.1** Comparison of steady state cold temperatures by numerical integration and FDM.

The steady state results predicted by present technique slightly overestimate those calculated by numerical integration. This error may be accounted for the following difference in two techniques. In numerical integration, all the calculations are done by assuming a certain mean temperature ($T_m$=255K in [16]). The properties of working gas and stack material are calculated at this particular mean temperature. In the present technique, the mean temperature is assumed to be equal to ambient the ambient temperature at start. The mean temperature is re-calculated after each iteration. The properties like the sound speed and specific heat capacity of stack material, which vary significantly with temperature, are also re-calculated at the updated mean temperature. A constant error of 10 K persists in case of Dr=3 %, while the maximum error in temperature calculation is 12 K at 1.5 % drive ratio. However, the qualitative trend in results of both techniques is very much similar.

## 5.2.2 Transient temperature profiles

The evolution of temperature profiles in the TAR which is designed in Chapter 3 is presented in this section. The working gas used is Helium. The stack material is taken as Mylar while the heat exchanger material is copper. The geometrical dimensions of the setup are same as those derived in Section 3.5.2. The effect of taper after the CHX and the buffer volume has been neglected while computing the temperature profile. The operating parameters are taken same as those used in the cooldown experiments. The resonance frequency of the TAR is experimentally found out to be 384 Hz which is close to the design frequency of 400 Hz. Similarly, the dynamic pressure measured downstream of the stack scales to about 5500 Pa at the driver surface. This scaling is done using eqn.(3.9). The operating parameters and the geometrical dimensions are listed in Table 5.1

**Table 5.1** Operating Parameters and Dimensions of various components

| Operating Parameters | | Stack (Mylar) Dimensions | | Resonator Dimensions | | Cold HX (Copper) Dimensions | |
|---|---|---|---|---|---|---|---|
| $p_m$ | 10 bar | $2l$ | 0.18 mm | L | 0.305 m | $L_s$ | 0.008 m |
| $f_{res}$ | 384 Hz | $2y_0$ | 0.30 mm | $A_{res}$ | 1.24E-4 m$^2$ | A | 8E-4 m$^2$ |
| $p_a$ | 5500 Pa | $L_s$ | 0.10 m | | | $x_s$ | 0.1517 m |
| $T_h$ | 308 K | A | 8E-4 m$^2$ | | | | |
| | | $x_s$ | 0.0977 m | | | | |

Figures 5.2-5.5 show the transient state temperature profiles in the TAR at various intervals of time after the driver is started at t=0. Initially at t=0, both the stack and the resonator are at ambient temperature. When the acoustic driver is switched on, there is a rapid formation of temperature gradient across the stack due to thermoacoustic enthalpy flow. The stack end closer to the driver (pressure antinode) begins to heat up while the other end (nearer to the pressure node) begins to cool down. Kink in the curves in Figure 5.2 shows the location of the cold heat exchanger of TAR.

Initially, it is also observed that the temperature of left end of the stack begins to rise above ambient even if the hot HX is maintained at ambient temperature. This is due to two reasons- firstly, during the transient regime of operation, the stack end closer to the driver (hot end) has to exchange much more heat with the ambient HX as compared to the cold end; secondly, the value of the heat transfer coefficient at the ambient HX location is lower than that of the cold

HX location. This is due to the fact that the cold HX is nearer to the velocity antinode than the ambient HX.

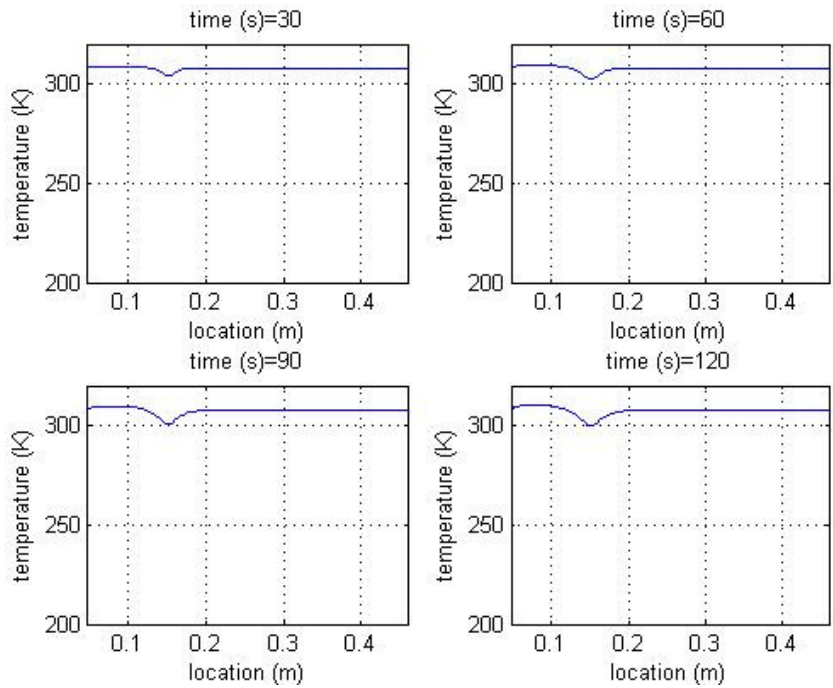

**Figure 5.2** Transient temperature profiles at various instants of time from t=30 s to t=120 s.

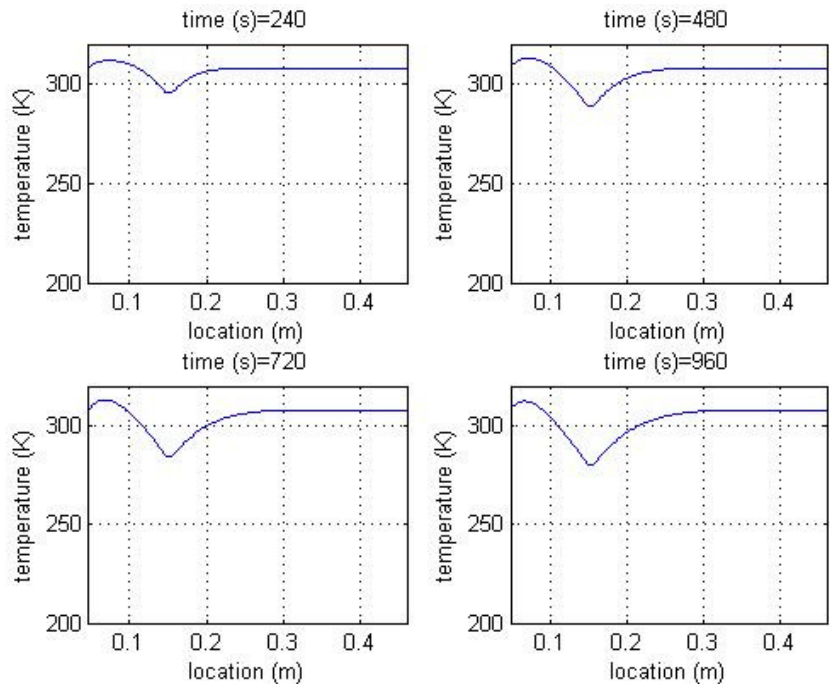

**Figure 5.3** Transient temperature profiles at various instants of time from t=240 s to t=960 s.

In the initial stages (Figures 5.2 & 5.3), there is hardly any noticeable fall in the resonator temperature because the heat diffusion process is much slower than the thermoacoustic heat pumping.

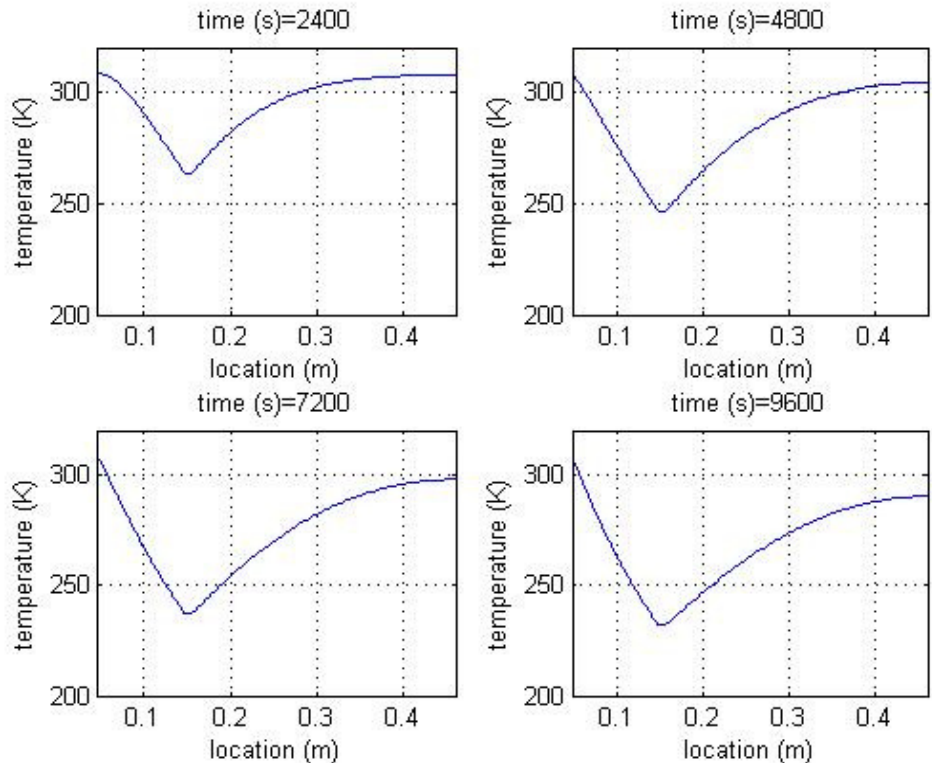

**Figure 5.4** Transient temperature profiles at various instants of time from t=2400s to t=9600s.

Figure 5.4 shows temperature profiles from t=2400 s to t=9600 s. After about t=2400 s the temperature of hot side of stack returns to ambient temperature following the boundary condition imposed at the ambient HX. This is because, the thermoacoustic heat flow decreases with the passage of time and hence, the ambient HX has less heat to transfer out. After approximately t=9600 s, a steep temperature gradient develops across the stack. At this time, the effect of heat diffusion can also be seen in the resonator. The entire resonator tube is below ambient temperature.

Figure 5.5 shows the temperature profiles from t=10800 s to t=21600 s. In this time interval, the temperature distribution in the stack has almost reached a steady state. The cold HX reaches its minimum temperature after approximately t=10800 s. At this point of time, the thermoacoustic heat pumping is counter balanced by the heat diffusing from the hot to the cold end of the stack. Hence, there is no net heat pumping from the cold to the hot end of the stack causing the cold HX temperature to stabilize.

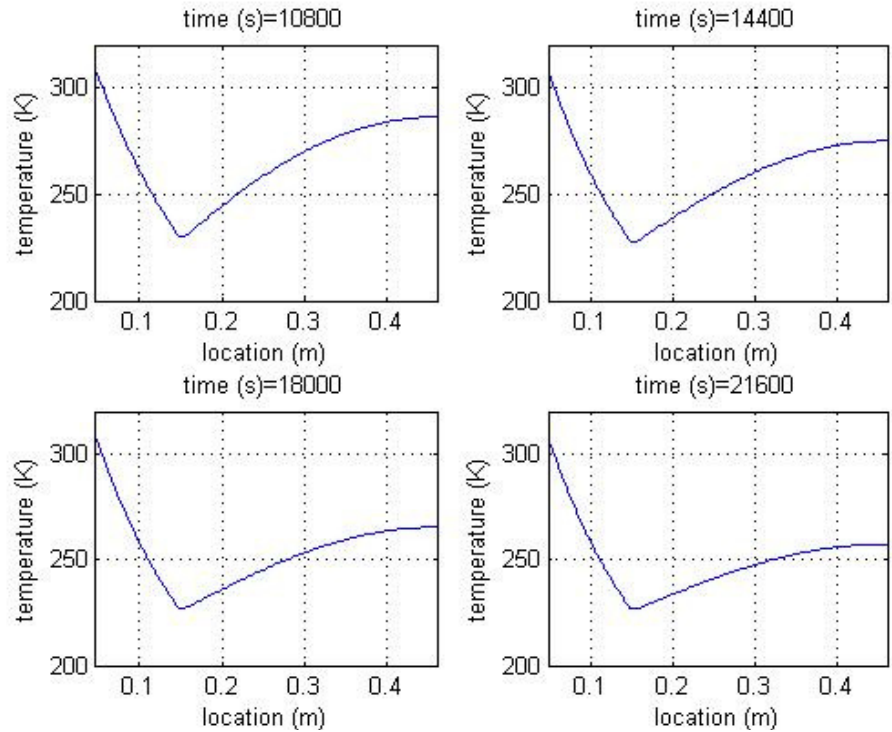

**Figure 5.5** Transient temperature profiles at various instants of time from t=10800 s to t=21600 s.

On careful inspection, it can also be seen in Figure 5.5, that after the cold HX cools to minimum, its temperature begins to rise slightly. This rise is due to diffusion of heat from the resonator to the cold HX. This temperature rise continues till the entire gas in hollow resonator cools down to near the cold HX temperature. In other words, the gas in the resonator acts as a 'heat load' on the cold HX, causing its temperature to rise. But the effect of this heat load is apparent only after the stack has reached a steady state.

The temperature profile at t=36000 s (10 hours) is shown in Figure 5.6. At this time, the average temperature inside the resonator is about to 230 K. This is close to the 227 K which is the CHX temperature at that time. Eventually, it is expected that the entire resonator would attain the CHX temperature at steady state.

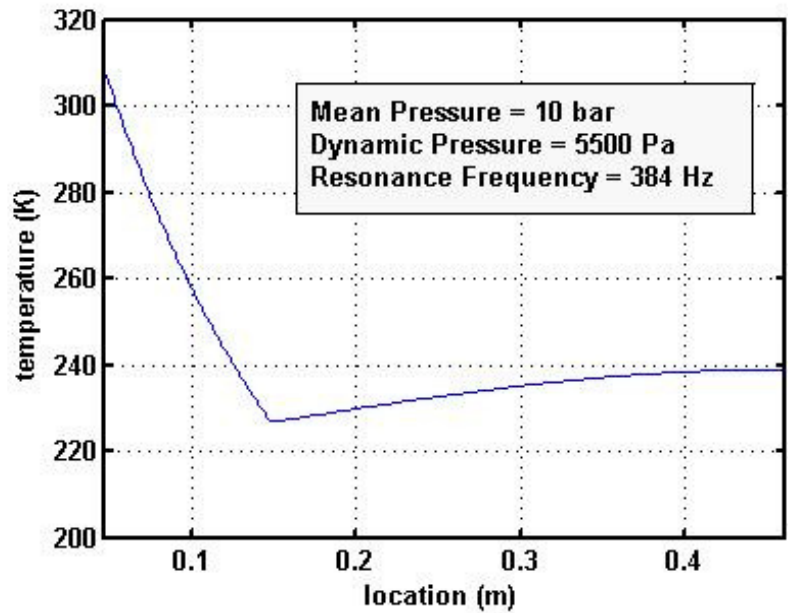

**Figure 5.6** Temperature Profile at t=36000 s. The CHX temperature is 227.2 K.

## 5.2.3 Cooldown Curve

The variation of temperature of cold HX with time from t=0 s to t=21600 s is given in Figure 5.7.

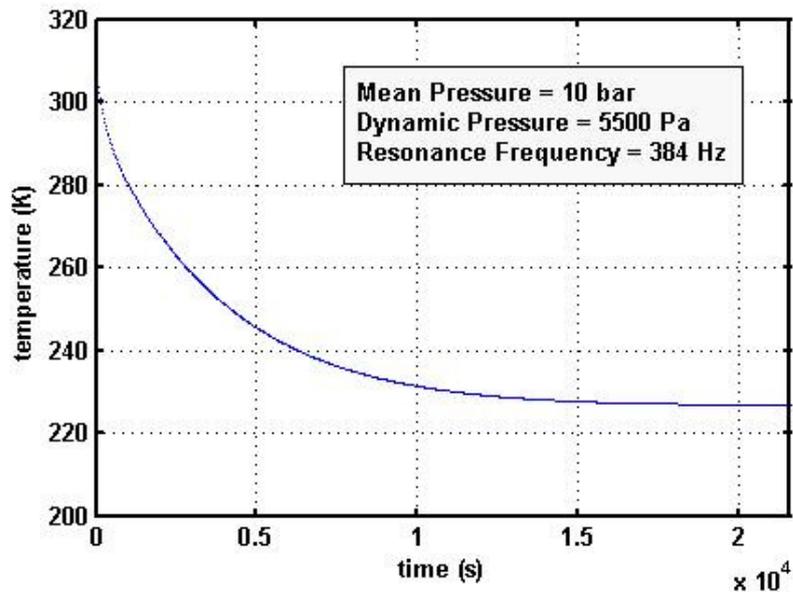

**Figure 5.7** Cooldown curve showing the variation of cold HX temperature with time.

It can be seen that in the initial stage of operation, the rate of fall of cold HX temperature is very fast. This is because $\Gamma \ll 1$, which causes very high thermoacoustic enthalpy flux to flow from cold end to hot end of the stack. Because of low temperature gradient across the stack, the reverse heat diffusion is also minimal in early stages of the operation. As the cold HX cools down with time, the temperature gradient builds up across the stack. As a result, the value of $\Gamma$ approaches 1 causing the magnitude of thermoacoustic enthalpy flux in the stack to fall down. At the same time, more heat starts to diffuse from the hot end to the cold end of stack. Hence, the cooldown rate becomes slower. The cold end of stack reaches a minimum value when the net heat flow through the stack becomes negligible. After the stack reaches steady state, the temperature of cold end begins to rise slightly. The reason for this has already been explained in the previous section.

The theoretical cooldown predicted by the present one-dimensional transient model is very rapid. This can be attributed to the following factors:

**a)** The present theoretical model does not take into account the lateral heat leaks into the TAR.

**b)** The cooldown of components like the stack holder and walls of resonator, outer body of cold HX; conduction losses through these components, etc is also not considered.

**c)** Due to heat exchanger ineffectiveness, the complete heat exchange limit over-estimates the values of heat transfer coefficients between the stack and the heat exchangers.

## 5.2.4 Cooldown Characteristics of Working Gases

This section presents the cooldown characteristics of the TAR when different working gases are used. As discussed above, the resonator takes much longer to cool down as compared to the stack and hence, only the cooldown of cold HX is presented and discussed. The operating parameters and the stack dimensions are taken same as given in Table 5.1 so that a comparison could be made for different working gases. Each gas has its own sound speed and so a given TAR will resonate at different frequencies for different gases. Instead, here the resonator length is adjusted so that the refrigerator always resonates at around 384 Hz for all the gases considered. The cooldown curves for different gases are shown in Figures 5.8a & 5.8b.

It can be seen from Figure 5.8a, that Helium has the fastest cooling rate in the transient regime. It is followed by Neon and then Argon. Neon cools to a minimum no load temperature among the three gases while the no load temperature of Argon is maximum.

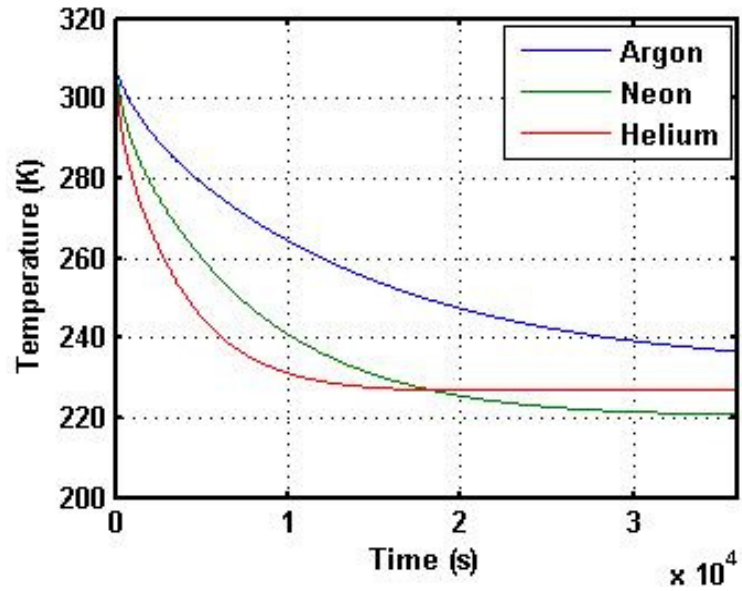

(a)

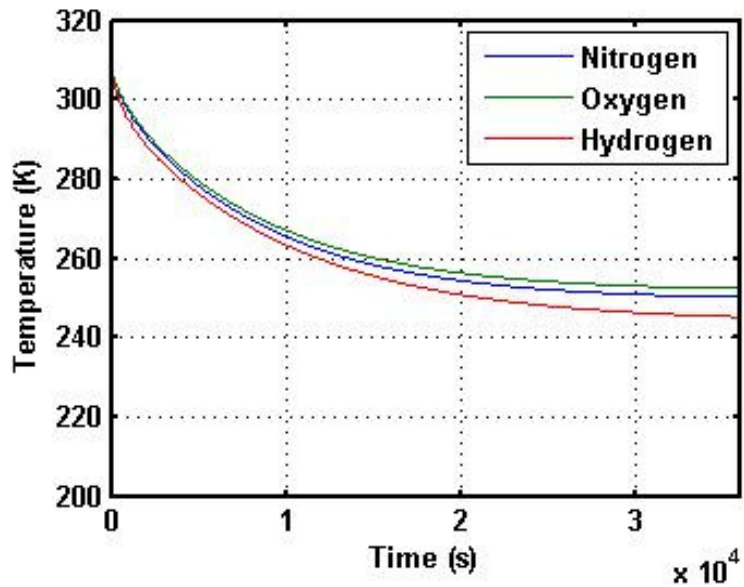

(b)

**Figure 5.8** Variation of cold HX temperature with time (a) monoatomic gases (b) diatomic gases.

All the gases cool to temperatures below 240 K. After reaching the minimum no load temperature, the temperature at cold HX begins to rise due to heat diffusion from the resonator. This is evident from the Helium curve. Though not very clearly visible in the figure, the other gases also follow the same trend. Similar observations can be made from Figure 5.8b, wherein Hydrogen has fastest cooldown rate and minimum no load temperature. It is followed by Nitrogen and then Oxygen. However, these di-atomic gases stabilize close to 250 K.

In the transient regime, the thermoacoustic heat pumping from cold to hot HX is much larger as compared to heat diffusion from hot to cold HX in stack. The rate of fall of temperature in stack depends on how fast heat is transported by the oscillatory velocity (thermoacoustic heat pumping) from one end of the stack to the other. It is evident from eqn.(3.10) that in standing wave refrigerators, the magnitude of oscillatory velocity is inversely proportional to '$\rho$a' product of working gas. Hence, as seen in Figures 5.8a & 5.8b, a gas with lower '$\rho$a' product has a faster cooldown rate.

The no load temperature that a gas can attain depends upon its volumetric heat capacity. For a certain amount of heat extracted from the gas, more the volumetric heat capacity less is the temperature drop. Furthermore, the volumetric heat capacity of a gas increases as its temperature falls. Hence with the passage of time, there is an increase in volumetric heat capacity of the gas. Also, the magnitude of thermoacoustic heat pumping (a measure of heat removed from gas) diminishes. Due to these two effects, a gas with higher volumetric heat capacity saturates to a higher steady state cold temperature.

Figures 5.8a & 5.8b also show that the cold end temperature of the stack begins to rise after reaching a minimum value. The reason for this rise is heat diffusion from gas in the resonator as already explained earlier. The rate of temperature rise is higher for gases having larger thermal diffusivity. Also, it can be concluded that a gas with higher thermal diffusivity will result in a faster cooldown of resonator.

The properties of gases viz. '$\rho$a', volumetric heat capacity '$\rho c_p$' and thermal diffusivity 'k/$\rho c_p$' calculated at 308 K and 10 bar mean pressure are shown below in Table 5.2.

Table 5.2 Properties of gases at 308 K, 10 bar

| Working Gas | '$\rho a$' (kg/m$^2$-s) | '$\rho c_p$' (J/m$^3$-K) | '$k/\rho c_p$' (m$^2$/s) (E-06) |
|---|---|---|---|
| Helium | 1641.25 | 8331.97 | 18.80 |
| Neon | 3658.81 | 8281.61 | 5.84 |
| Argon | 5187.07 | 8537.41 | 2.10 |
| Hydrogen | 1059.26 | 11473.02 | 11.30 |
| Nitrogen | 3981.09 | 11853.58 | 4.19 |
| Oxygen | 4233.82 | 11983.00 | 4.22 |

## 5.3 Dynamic Pressure Measurements

### 5.3.1 Driver Parameters

In the course of adapting the commercial speaker to the purpose of driving the TAR, the mechanical parameters of the driver get modified. Therefore in the present work, these are determined experimentally. The voice coil, the cone and the suspension are accurately weighed to find the approximate new moving mass, $M_m$. The new suspension stiffness, $k_m$ is determined by measuring the resonance frequency in vacuum. The force factor, Bl is calculated by measuring the magnetic gap flux and the active coil length. The electrical parameters of the coil, *viz.* $R_e$ and $L_e$ remain unchanged during modification of the speaker. The mechanical resistance, $R_m$ is calculated from the measured value of $Z_{et}$ at resonance in vacuum, Bl and $R_e$. The electrical reactance being small compared to $R_e$, is neglected while calculating $R_m$. The driver parameters in the present case (with the old voice coil) are given in Table 5.3.

Table 5.3 Drive Parameters

| Parameter | Value |
|---|---|
| $R_e$ | 7.2 Ω |
| $L_e$ | 1.03 mH |
| Bl | 10.0 T-m |
| $M_m$ | 0.026 kg |
| $k_m$ | 7000 N/m |
| $R_m$ | 4.58 N-s/m |
| S | 0.0008 m$^2$ |
| $f_{res}$ in vacuum | 78 Hz |

## 5.3.2 Resonance Frequency of Half Wave Resonators

The resonance of an electrical system having a complex impedance, is characterized by its minimum input current. The frequency at which the current drawn by the system is minimum, is defined as the resonance frequency of the system. Figure 5.9 shows the current drawn by the driver-resonator assembly over a frequency range when the input voltage is kept constant at 5 V. The measurement in vacuum ($10^{-3}$ mbar) shows only one minimum of current, where the system is resonant (driver mechanical resonance). When the system is charged with Helium or Nitrogen at 9.5 bar, two current minima are observed. The driver resonance frequency still remains close to 85 Hz as in vacuum, while the acoustic load (charged gas column) resonates close to 400 Hz as predicted by eqn(3.5).

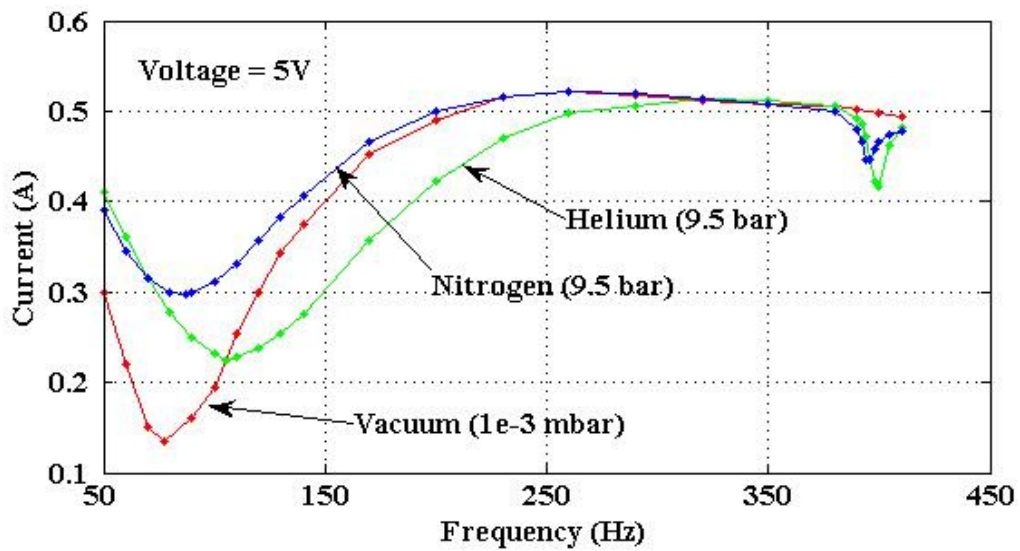

**Figure 5.9** Variation of current with operating frequency

The driver resonance frequency still remains close to 85 Hz as in vacuum, while the acoustic load (charged gas column) resonates close to 400 Hz as predicted by eqn(3.5). A slight rise in the driver resonance frequency is observed due to increased stiffness of driver when the gas is charged at high pressure.

## 5.3.3 Effect of Operating Frequency on Dynamic Pressure

In standing wave TAR systems, the phasing between the dynamic pressure and oscillatory velocity is very crucial for production of low temperatures. The proper phasing can be achieved by operating the system near resonance frequency of the acoustic load. Furthermore, when the acoustic load resonates, the dynamic pressure obtained is also large. This can be seen from Figure 5.10. It shows the dynamic pressure as the operating frequency is varied around the resonance at a constant supply voltage of 5 V. In case of both, the theoretical prediction and the experiments, dynamic pressure increases as the operating frequency approaches the load resonance frequency. It exhibits a maxima peak near the resonance frequency and again starts to diminish after the resonance cross-over.

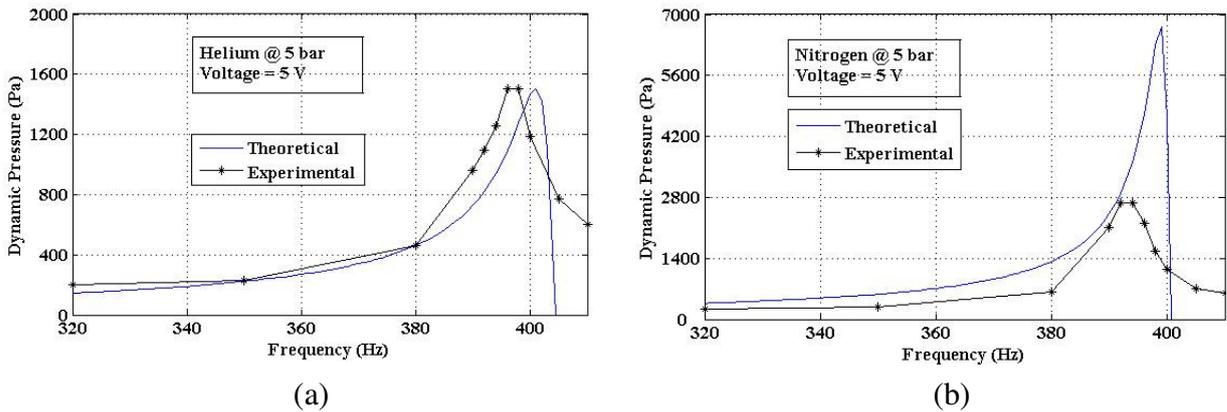

(a)                                        (b)

**Figure 5.10** The dynamic pressure in the resonator as a function of operating frequency
(a) with Helium; (b) with Nitrogen.

A comparison between theory and experiment shows a very good agreement in case of Helium (Figure 5.10a). However, in case of Nitrogen as seen in Figure 5.10b, very high dynamic pressure is predicted around resonance as compared to that observed during experiments. The deviation is attributed to the non-linear effects explained in a later section.

### 5.3.4 Effect of Charging Pressure and Working Gas on Dynamic Pressure

The density of working gas which appears in the expression of acoustic impedance (eqn(3.4)), is one of the parameters that governs the dynamic pressure in a TAR. The density of a gas is proportional to its pressure and it is more intuitive to study this dynamic pressure dependency in terms of the charging pressure of the gas. Figure 5.11 shows a comparison between the theoretically predicted and experimentally obtained dynamic pressure at resonance for several input voltage levels. The test is carried out for Helium and Nitrogen at three different values of changing pressure *viz.* 5 bar, 8 bar and 9.5 bar. In case of both the gases, the dynamic pressure for a given value of input voltage increases with the charging pressure. This can be interpreted in following manner. In an electric circuit where the components are in series, the voltage drop across a certain component will increase if its impedance is increased. In a similar fashion, the dynamic pressure, which is analogous to the 'voltage drop' across the acoustic impedance, should increase with increase in the acoustic impedance. This is predicted theoretically as well as observed in experiments. Furthermore, it can be seen that for a certain value of charging pressure and voltage, Nitrogen induces larger dynamic pressures. This is because Nitrogen being heavier that Helium, has more density at same pressure and temperature. As a result, Nitrogen offers greater impedance to the driver.

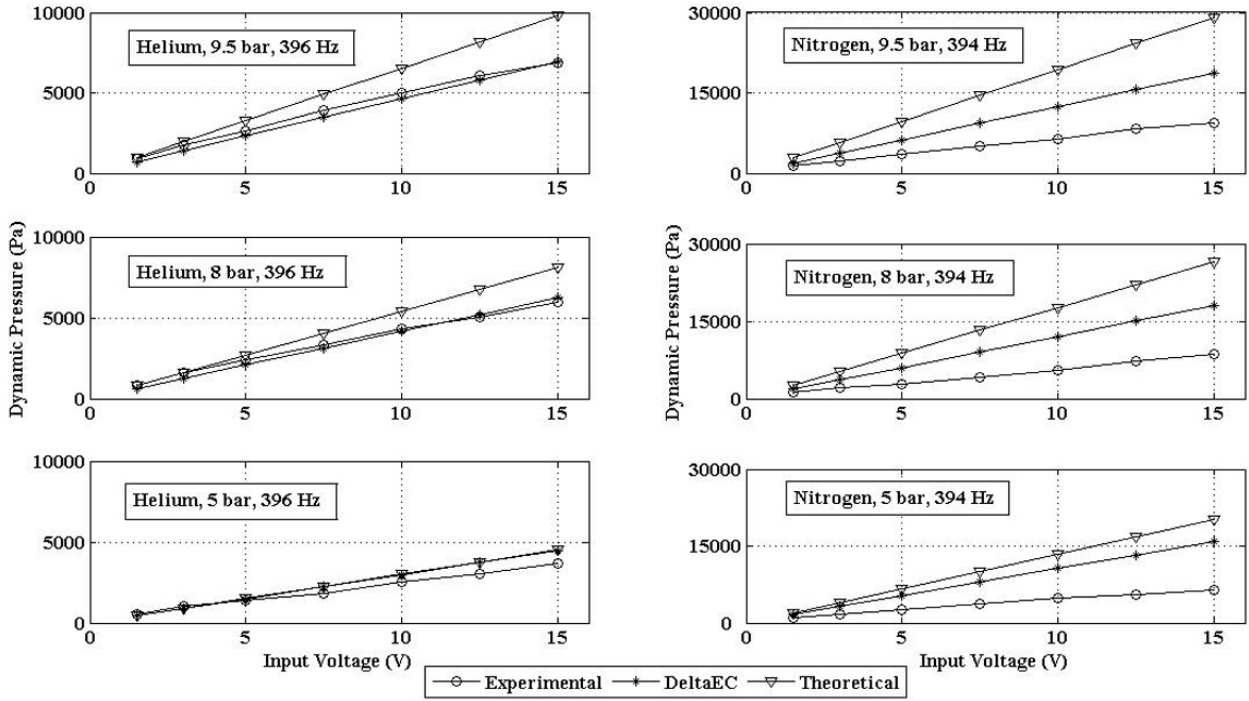

(a) (b)

**Figure 5.11** The dynamic pressure near the driver piston as a function of input voltage at various charging pressures when operated at resonance **a)** with Helium **b)** with Nitrogen as working gas.

## 5.3.5 Non-Linear Effects

Referring to Figure 5.11, there is a fair match between theoretical and experimental values of dynamic pressure when the input voltage is small. When the voltage is increased, the dynamic pressure is expected to increase linearly as predicted by the theory. However, this is very unlike what is observed from experiments. Further, at large voltages, the deviation from theory is quite large. This deviation can be accounted by the non-linear effects that are prevalent in resonators with uniform cross sections [18] when driven at large voltage excitation. Straight resonators with uniform cross sections have resonance modes that are integral multiple of the fundamental mode. In non linear regime, the acoustic energy which the driver pumps into the acoustic load, is transferred to higher harmonics of the fundamental. In this way, the energy of the fundamental mode is lesser than it could have been in case of linear regime. One such non linear effect observed in our experiments was the phenomenon of periodic shocks [29]. Such shocks exist when a straight gas column is excited with large amplitudes of the piston near its resonance frequency. The images in Figure 5.12 show the waves observed on oscilloscope which represent the dynamic pressure in half wavelength Nitrogen resonator charged at 9.5 bar pressure.

At all the three voltage levels (5 V, 10 V and 15 V), the dynamic pressure wave is very much sinusoidal when the system is operated slightly away from resonance (390 and 400 Hz). Very close to resonance, periodic shocks set in and disturb the perfect sinusoidal nature of the waveform. It can also be seen that the shocks are more pronounced at higher voltage levels. Hence, the deviation of experiments from theory is large at higher level of input voltages.

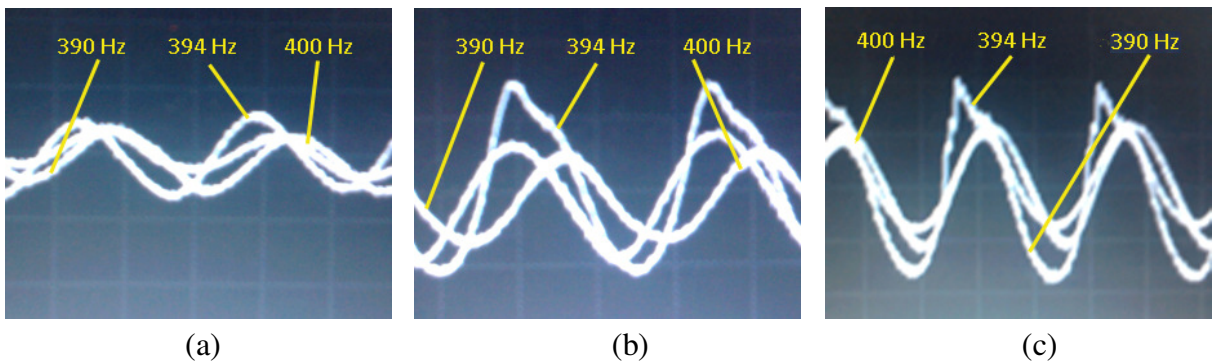

(a)　　　　　　　　　　　　(b)　　　　　　　　　　　　(c)

**Figure 5.12** The waveforms observed on oscilloscope representing dynamic pressure in Nitrogen resonator at voltage levels- (a) 5 V; (b) 10 V; (c) 15 V.

To further account for the non-linear effects, a DeltaEC model of the setup is made and the simulations for the tests described in Figure 5.11 are carried out. The dynamic pressure predicted by DeltaEC in case of Helium follows very closely with the experimental findings. Deviations are still observed in case of Nitrogen mainly due to existence of periodic shocks in the resonator. However, the DeltaEC predictions follow the experimental results more closely as compared to those given by the impedance transfer technique.

## 5.4 Experimental Results

Experiments are carried out with the TAR setup in order to determine the cold temperature that can be attained. The quarter wave resonator along with the buffer volume is used in these experiments. The details of experimental setup and instrumentation have already been explained in the previous chapter.

At first, the resonance frequency of the quarter wave resonator is determined by giving a frequency sweep. At resonance, a local minima of current is observed on the digital power meter. This is accompanied with an occurrence of maxima of the dynamic pressure waveform as seen on the digital oscilloscope. The experimentally observed resonance frequency is 384 Hz. This value matches closely with the design frequency of 400 Hz. Similar tests are done at different charging pressures so as to confirm that the resonance frequency is independent of the gas pressure.

For the cooldown measurements, the TAR is run at the resonance frequency and 20 W of electrical input power. This power input is restricted by the current limit of the coil which is 2.5 A. To study the effect of charging pressure on cooldown, the tests are carried out at different charging pressures *viz.* 4 bar, 6 bar, 8 bar and 10 bar. To prevent unnecessary heating of the acoustic driver, each test is run for a duration of 20 minutes. After each run, the power is switch off so that the cold end warms up to ambient temperature.

The cooldown curves for different charging pressures are shown in Figure 5.13

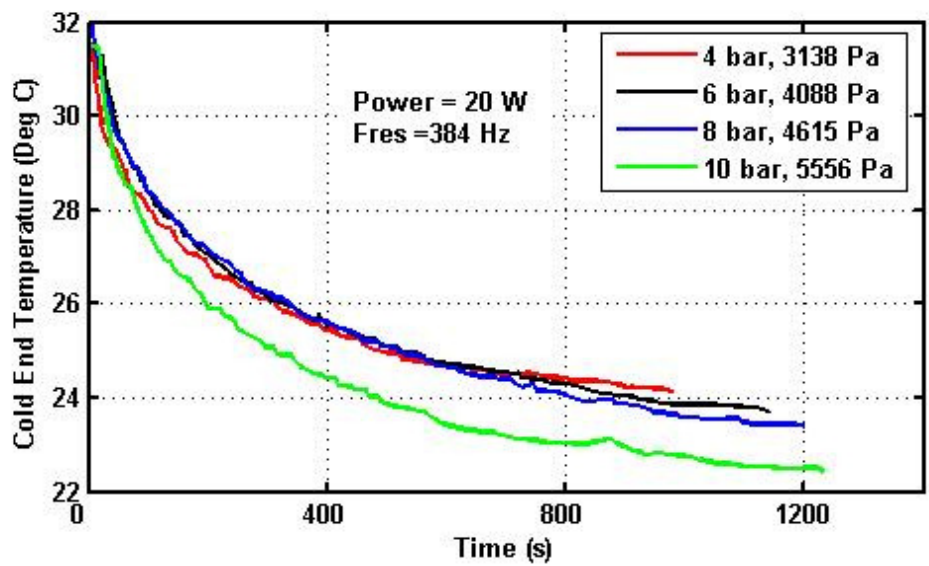

**Figure 5.13** Cooldown curves for different charging pressures

It is seen from Figure 5.13 that after a rapid fall, the cold end temperature stabilizes after approximately 20 minutes after switching on the power. The effect of charging pressure can also be seen. The cold end temperature is lower for higher value of charging pressure.

As discussed in earlier sections, the no load temperature at the cold end of the stack depends on the dynamic pressure inside the TAR. It is also seen from the results of section 5.3.2, that the dynamic pressure induced in the TAR itself depends on the charging pressure. This behavior of dynamic pressure can be seen from Figure 5.14. It shows the dynamic pressure measured downstream of the stack during the cooldown measurements.

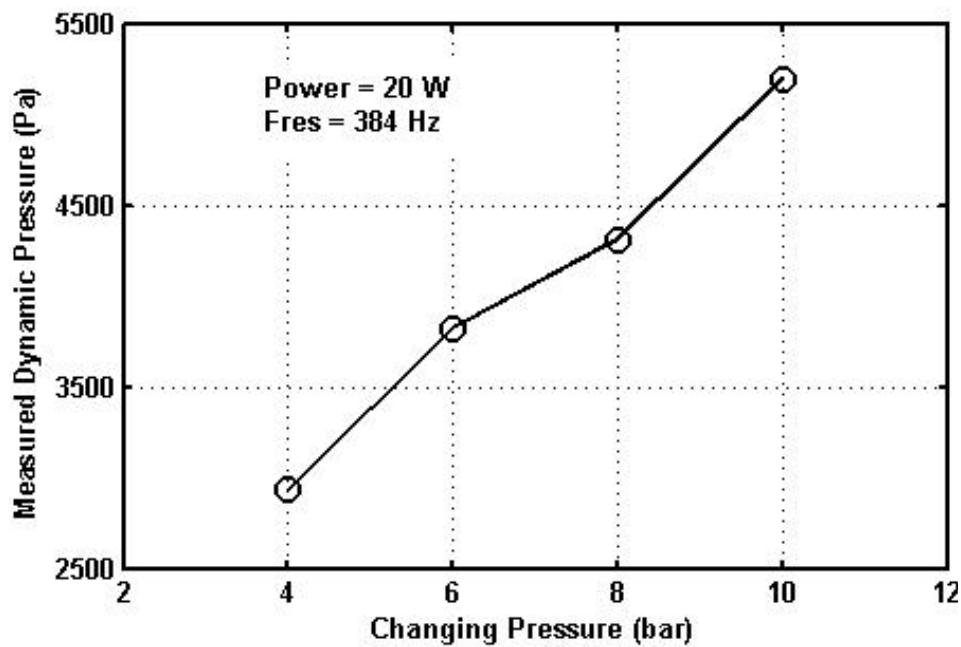

**Figure 5.14** Dynamic Pressure during cooldown measurements

Figure 5.15 shows the steady state cold end temperatures for various charging pressures. The temperature lift at the cold end defined as $T_{ambient} - T_{cold}$ is also shown in the figure. The minimum cold end temperature of **22.27 °C (~ 295 K)** and the corresponding temperature lift of **9.86 °C** have been observed in the experiments.

The steady state cold end temperature predicted by the transient state model for the above developed TAR configuration is 227 K. Temperatures of such nature have been reported in literature [3,16].

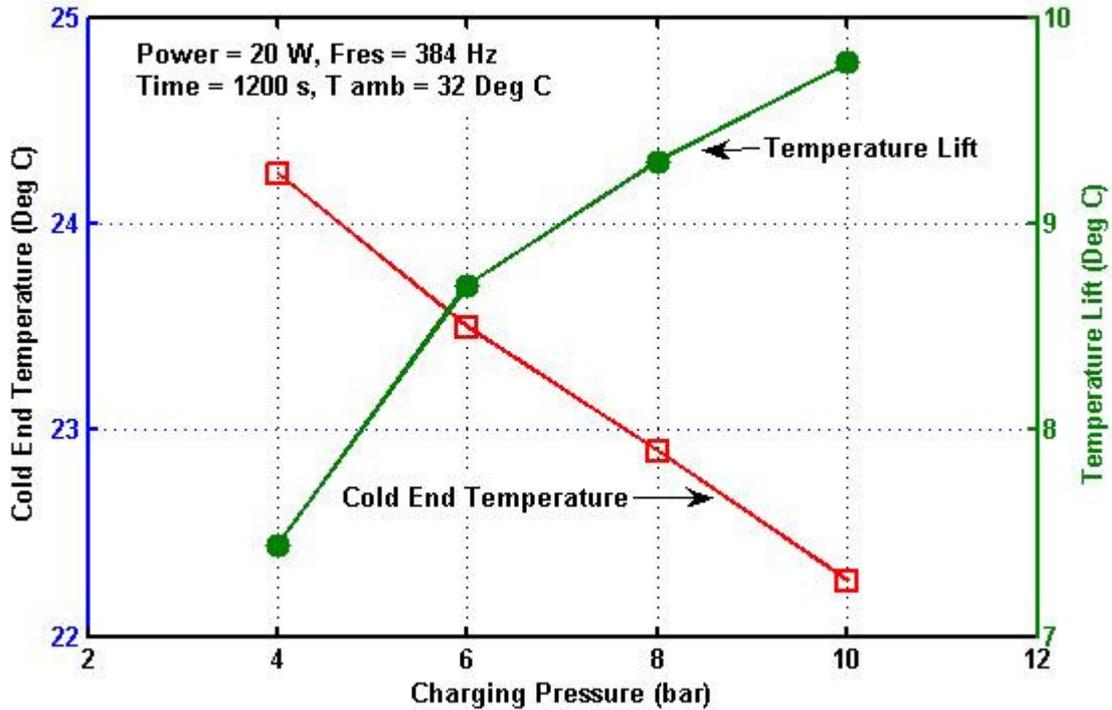

**Figure 5.15** Variation of steady state cold end temperature and temperature lift for various charging pressures.

However, the lowest experimentally observed cold end temperature for the same operating parameters is close to 295 K, which is just about 10 K below ambient. Thus, there is a huge deviation of experimental observations from theory. This deviation can be explained with the help of following:

a) **Lack of insulation around the cold HX.**

In the present TAR assembly, cold temperature measurements are done without an insulation or a vacuum around the assembly. As a result, there is unwanted heat load due to heat leak from the ambient to the cold HX block (though the numerical value of this heat leak has not been determined).

The maximum dynamic pressure obtained inside the TAR setup 5550 Pa. This is very less as compared to values like 30000 Pa as reported by Holfer [16]. At such low dynamic pressure, even a small heat load at the cold end of the stack has a drastic effect on the cold end temperature. To justify this fact, a DeltaEC model of Hofler's TAR is created and is run for different values of dynamic pressures ranging from 30000 Pa to 8000 Pa. The plots are shown in Figure 5.20. The curves corresponding to 30000 Pa and 15000 Pa match closely with those given in Figure 5.5 or [1,16]. This validates the DeltaEC model. The third curve corresponds to a dynamic pressure of 8000 Pa. As seen from the figure, when the dynamic pressure is high, the temperature ratio *vs* the heat load line is flat. As the dynamic pressure is lowered, the slope of the line increases and the line becomes steeper. To conclude, as the dynamic pressure is lowered, the cold end temperature becomes highly sensitive to the heat load at the cold HX.

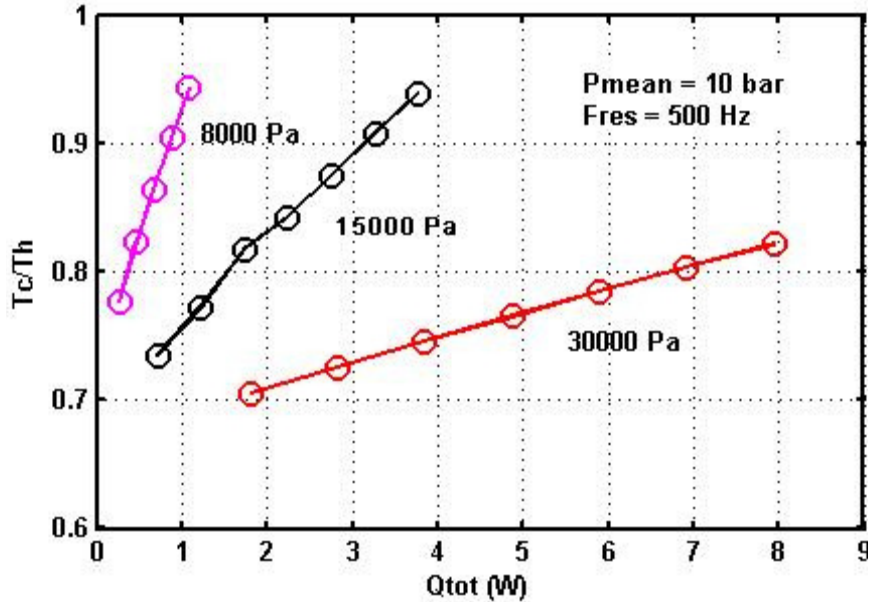

**Figure 5.16** Response of cold end temperature to the variation of heat load. DeltaEC model of Hofler's TAR

**b) Ambient HX**

The thermoacoustic effect causes heat to flow from one end of the stack to the other. As, a result, the stack end near to the pressure antinode begins to heat up while that near to the pressure starts to cool down. Thus, a temperature difference ($T_h - T_c$) is generated across the stack. The cold temperature $T_c$ becomes low only when the level of high temperature $T_h$ is maintained close to ambient. This has to be accomplished by removing heat from the stack end which is getting heated up. For this purpose, a proper design of heat exchanger is necessary. In present course of work, less emphasis was given on design of heat exchangers and hence, the ambient heat exchanger was not efficient. As a result, the hot end of stack was heated well above the ambient temperature (as high as 45 °C when ambient is at 32 °C). Thus, the assumed boundary condition of isothermal ambient HX in the transient model was not realized during the experiments.

**c) Resistive Heating in the Acoustic Driver**

The cold end temperature of 227 K is predicted by the transient model of the TAR. However, it should also be noted that, the TAR will theoretically take 21600 s (6 hours) of continuous operation to cool to this temperature (Figure 5.11). However, such a long run may damage the acoustic driver due to resistive heating of the voice coil. To prevent this, an arrangement to remove this heat from the driver has to be done. As mentioned in literature [3], this can be done by brazing cooling water tubes of copper around the acoustic driver chamber with a high flow-rate of cooling water. In present case, due to time constraints, such a mechanism couldn't be implemented. Hence, there was a restriction on the duration of each experimental run. The run-time for each trial was deliberately kept to 20 minutes.

# Chapter 6
# CONCLUSIONS AND FUTURE SCOPE

## 6.1 Conclusions

In-depth theoretical investigations of Standing Wave TARs have been done in the course of present work. Various theoretical TAR design models reported in literature have been studied. A theoretical model has been developed to study the transient state of a TAR. This model is based on the Linear Theory of Thermoacoustics. Using this model, the transient temperature profiles and the cooldown characteristics of a given TAR configuration have been predicted. The model has been further extended to a theoretical study the cooldown characteristics of various working fluids.

The phenomenon of standing wave resonance has been well understood through present study. A parametric study of dynamic pressure in a loudspeaker driven TAR has been done both theoretically and experimentally. The charging pressure and operating frequency are used as the independent parameters.

The design aspects of TAR- the choice of operating parameters, the geometric dimensions, the choice of materials and the fabrication techniques have been explored. A TAR driven by a commercially available moving coil loudspeaker has been designed and constructed. It employs readily available metallic and delrin tubings for its construction. As such, no sophisticated manufacturing techniques were required for fabrication.

Operating with Helium gas charged at 10 bar and consuming 20 W of electrical power, the refrigerator reached a minimum cold temperature of **22.27 $^o$C** with a temperature lift of **9.86 $^o$C** from the ambient.

Following are some of the important conclusions drawn from the present work:-

**1)** Refrigerators based on thermoacoustic technology are capable of reaching low temperatures at the cost of quite meagre acoustic power. However, their overall efficiency is hampered when they are driven by acoustic drivers. This is because the electro-acoustic efficiency of drivers is very low. The typical range of electro-acoustic efficiency of drivers specially designed for thermoacoustic refrigeration is 25-35 % [22,28].

**2)** From the transient state analysis it is observed that, the cooldown time of resonator is several times that of the stack. For faster cooldown, the resonator length should be kept as low as possible.

**3)** The no-load temperature that a working gas can attain depends upon its volumetric heat capacity. Gases with lesser volumetric heat capacity cool to lower temperatures.

**4)** A lighter gas has a higher cooldown rate because it has a higher oscillatory velocity for a given dynamic pressure amplitude.

**5)** The impedance transfer technique accurately determines the resonance frequency of an acoustic load. The dynamic pressure prediction at lower level of excitation is also very good. However, the technique fails at high levels of excitation due to non-linear effects.

**6)** Resonance frequency plays a vital role in the operation of a TAR. It is found that the dynamic pressure attains a maximum value when the operating frequency is close to the resonance frequency. Away from resonance, the dynamic pressure is negligible even at high input power.

**7)** Dynamic pressure in a given resonator column can be made to increase by increasing the load impedance. This can be done by increasing the charging pressure or by using a denser working gas.

**8)** Although easy to design and fabricate, straight resonators are not suitable for Standing Wave TARs. This is because of the non-linear effects that are prevalent in straight resonators driven at resonance frequency and high amplitudes. Periodic shock waves have been observed in the experiments. These non-linear effects hamper the amplitude as well as phasing of the standing wave.

**9)** In order to cool to low temperatures, high dynamic pressures are required. At low dynamic pressures, the cooldown rate of a TAR is very slow and the cold end is highly sensitive to external heat leaks.

## 6.2 Future Scope

A realistic validation of the transient model by the method of experiment could not be achieved during the present investigations. In view of this, following are some of the areas which need to be investigated on, in the course of future work:

**1) Transient State Model**

The effect of TAR hardware on the cooldown characteristics of a TAR needs to be incorporated into the present transient state model. This will include taking into consideration the cooldown of stack holder, heat exchangers and the resonator walls. The heat transfer coefficients for oscillatory flows also need to be modeled more realistically.

**2) Acoustic Driver Efficiency**

The efficiency of the acoustic driver can be maximized by making it resonate at the design frequency of the TAR. In the present study, a commercial loudspeaker motor was used and hence, mechanical resonance of the driver could not be achieved. This caused the driver to draw more current to generate a certain dynamic pressure. As a result, the current limit of the voice coil reached at very low dynamic pressure levels. In the present work, a very light voice coil with a total mass of 12 gm has been successfully manufactured. However, the stiffness of the rubber suspension still remained too low to shift up the driver mechanical resonance near the operating frequency. Hence, proper suspensions with high enough stiffness need to be designed. The use of gas-spring as suggested by Tijani et.al [22] can also be implemented after proper analysis.

In order to construct a powerful driver, magnetic materials with higher strength and pole piece material with higher permeability are needed. Along with this, the arrangement of the magnet and pole pieces is to be designed so as to maximize the field strength in the magnetic flux gap. The length of the voice coil has to be minimized so as to decrease the DC resistance and the weight. To decrease the weight, super-insulated CCA (Copper Clad Aluminum) wires may be used, but at the expense of increased resistance. The two opposing effects can be countered by proper optimization. The uniform suspension of voice coil in the magnetic flux gap is also important in order to minimize the damping or the mechanical resistance.

**3) Designe of Heat Exchangers**

The heat exchangers are needed to carry heat in and out from the refrigerator. In the present work, the design of heat exchangers for oscillating flows hasn't been studied extensively. The heat exchangers have been fabricated based on the fact that their optimal length is equal to twice the local acoustic displacement of gas. However, the mechanisms of heat transfer from the gas to the HX matrix inside the TAR and the techniques to transfer this heat outside the refrigerator need special focus.

**4) Fabrication of Stack**

The stack is called 'the heart' of a TAR. Its proper design and fabrication is of utmost importance. Though many design methodologies are available [1,7,18], these are based on the 'short stack' assumption that the stack does not hamper the standing wave. However, as observed in present work, an improper fabricated stack causes the dynamic pressure to decrease tremendously. Hence, the fabrication aspects of the stack need to be studied. The use of other geometries of stack is also viable. The stack can be constructed using other materials like ceramic, kapton, etc.

**5) Quarter Wave Resonators**

As suggested in Section 3.3, an optimization algorithm may be developed to design a resonator that would yield high amplitudes of dynamic pressure at the design frequency. Such resonators with non-uniform cross section are void of non-linear effects.

# Publications

1. R. C. Dhuley, M. D. Atrey, *"Design guidelines for a thermoacoustic refrigerator"*, Indian Journal of Cryogenics, 35(1-4), 2010.

2. R. C. Dhuley, M. D. Atrey, *"Investigations on a Standing Wave Thermoacoustic Refrigerator",* paper presented at the 16$^{th}$ International Cryocooler Conference at Atlanta, GA, May 2010.

3. R. C. Dhuley, M. D. Atrey, *"Transient Analysis of a Thermoacoustic Refrigerator"*, manuscript under preparation.

# Acknowledgement

I whole heartedly thank my guide, Prof. Milind D. Atrey for introducing me to *'Thermoacoustics'* at a very early stage of my IITB curriculum. It is his insights and motivation that have helped my interest in the subject to blossom. I sincerely thank him for his understanding, guidance and support during the course of my Dual Degree Project, which has been an immense learning experience.

The help of Mr. Vasudev Khandare and Mr. Pandurang Bhandare during the course of fabrication and experimentation is remarkable. Without their genius, many of my mistakes would have simply been irreparable. I also acknowledge the help and support of Mrs. Shweta Mane without whom, maintaining the project finances would have been unimaginable.

It is with deep sense of gratitude that I acknowledge the support and help of Mr. Rohit Mehta, Mr. Kedar Sant, Mr. Purushottam Ardhapurkar and Mr. Amarish Badgujar. The several technical and non-technical discussions with Mr. Mandar Tendolkar and Mr. Jagan Mohan as well as their help in my experimentation, are invaluable to me. I also thank Mr. Suraj Abdan, Mr. Rajeev Hatwar and the two interns, Puneet and Mrinal for making the learning environment joyous.

I also thank Mr. Madhukar Bhandurge of Mumbai Repowering Company for his apt help in fabrication. The timely support of Mr. Sunit Subhedar, Mr. Milind Patil and Mr. Damodar Ratha of Peerless Audio for the loudspeakers, and Mr. Aseem Mehta of Lamtex Insulations for the Mylar film is quite noteworthy.

I thank all the staff of the Refrigeration and Cryogenics Lab as well as the Liquid Nitrogen Plant for their help and support.